\documentclass[aps,amsmath,amssymb,twocolumn,nofootinbib]{revtex4-1}
\usepackage{amsmath,amssymb}
\usepackage{graphicx}
\usepackage{multirow}
   \usepackage{physics} 
\usepackage{acronym}
\usepackage{hyperref}
\usepackage{color}
 
\newcommand{\beq}{\begin{eqnarray}}
\newcommand{\eeq}{\end{eqnarray}}

\begin{document}

\title{\emph{You better watch out}: US COVID-19 wave dynamics versus vaccination strategy }

\author{Giacomo Cacciapaglia} 
\email{g.cacciapaglia@ipnl.in2p3.fr}
\affiliation{\mbox{Institut de Physique des deux Infinis de Lyon (IP2I),  UMR5822, CNRS/IN2P3, F-69622, Villeurbanne, France}}
\affiliation{\mbox{University of Lyon, Universit{\' e} Claude Bernard Lyon 1,  F-69001, Lyon, France}}

\author{Corentin Cot} 
\email{cot@ipnl.in2p3.fr}
\affiliation{\mbox{Institut de Physique des deux Infinis de Lyon (IP2I),  UMR5822, CNRS/IN2P3, F-69622, Villeurbanne, France}}
\affiliation{\mbox{University of Lyon, Universit{\' e} Claude Bernard Lyon 1,  F-69001, Lyon, France}}

  \author{Anna Sigridur Islind}
\email{islind@ru.is}
\affiliation{\mbox{Department of Computer Science, Reykjavík University, Menntavegur 1, 102 Reykjavík, Iceland}}

  \author{María \'Oskarsd\'ottir}
\email{mariaoskars@ru.is}
\affiliation{\mbox{Department of Computer Science, Reykjavík University, Menntavegur 1, 102 Reykjavík, Iceland}}

   \author{Francesco Sannino}
\email{sannino@cp3.sdu.dk}
\affiliation{CP3-Origins \& the Danish Institute for Advanced Study, University of Southern Denmark, Campusvej 55, DK-5230 Odense, Denmark}
\affiliation{ Dipartimento di Fisica E. Pancini, Universit\`a di Napoli Federico II \& INFN sezione di Napoli, Complesso Universitario di Monte S. Angelo Edificio 6, via Cintia, 80126 Napoli, Italy.}

\vspace{0.5cm}
\begin{abstract}
\textbf{Abstract:} We employ the epidemic Renormalization Group (eRG) framework to understand, reproduce and predict the COVID-19 pandemic diffusion across the US. The human mobility across different geographical US divisions is modelled via open source flight data alongside the impact of social distancing for each such division. We analyse the impact of the vaccination strategy on the current pandemic wave dynamics in the US. 
We observe that the ongoing vaccination campaign will not impact the current pandemic wave and therefore strict social distancing measures must still be enacted. 
To curb the current and the next waves our results indisputably show that vaccinations alone are not enough and strict social distancing measures are required until sufficient immunity is achieved.  
Our results are essential for a successful vaccination strategy in the US.
  \end{abstract}

\maketitle

\section{Introduction}
The United States (US), raged by the SARS-CoV-2 virus, are paying an immense toll in terms of the loss of human lives and jobs, with a dreadful impact on society and economy. 
Understanding and predicting the time evolution of the pandemic plays a key role in defining prevention and control strategies. 
Short-term forecasts have been obtained, since the early days, via effective methods \cite{Perc2020,Hancean2020,Zhou2020}. Furthermore, time-honored mathematical models can be used, like compartmental models \cite{SEIR,scala2020,friston2020second,sonnino2020stochastic,Anas2020} of the SIR type \cite{Kermack:1927} or complex networks \cite{ZHAN2018437,PERC20171,WANG20151}.
Nevertheless, it remains very hard to understand and forecast the wave pattern of pandemics like COVID-19 \cite{Scudellari}.


In this work, we employ the \emph{epidemic Renormalization Group} (eRG) framework, recently developed in \cite{DellaMorte:2020wlc,Cacciapaglia:2020mjf}. It can be mapped \cite{Cacciapaglia:2020mjf,DellaMorte:2020qry} into a time-dependent compartmental model of the SIR type  \cite{Kermack:1927}. The eRG framework provides a single first order differential equation, apt to describing the time-evolution of the cumulative number of infected cases in an isolated region \cite{DellaMorte:2020wlc}. It has been extended in~\cite{Cacciapaglia:2020mjf} to include interactions among multiple regions of the world.
The main advantage over SIR models is its simplicity, and the fact that it relies on symmetries of the system instead of a detailed description. As a result, no computer simulation is needed in order to understand the time-evolution of the epidemic even at large scales \cite{Cacciapaglia:2020mjf}.
Recently, the framework has been extended to include the multi-wave pattern \cite{cacciapaglia2020evidence,cacciapaglia2020multiwave} observed in the COVID-19 and other pandemics \cite{1918influenza}.

The Renormalization Group approach \cite{Wilson:1971bg,Wilson:1971dh} has a long history in physics with impact from particle to condensed matter physics and beyond. Its application to epidemic dynamics is complementary to other approaches \cite{LI2019566,ZHAN2018437,PERC20171, WANG20151,WANG20161,Danby85,Brauer2019,Miller2012,Murray,Fishman2014,Pell2018}.  
Here we demonstrate that the framework is able to reproduce and predict the pandemic diffusion in the US taking into account the human mobility across different geographical US divisions, as well as the impact of social distancing within each one.
To gain an insight and to better monitor the human exchange we make use of open source flight data among the states. We calibrate the model on the first wave pandemic, raging from March to August, 2020. With these insights, we then analyse and understand the current second wave, raging in all the divisions.
The eRG framework can also be easily adapted to take into account vaccinations \cite{WANG20161}. We propose a new framework and use it to quantify the impact of the vaccination campaign, started on December 14th, on the current and future wave dynamics. 
Our results are in agreement with previous work based on compartmental models \cite{vaccine}, and confirm that the current campaign will have limited impact on the ongoing wave.

\section{Methodology}
In this section we briefly review our methods that include the open source flight data description, their interplay with the eRG mathematical model framework and, last but not least, the interplay with vaccine deployment and implementation. 

\subsection{Data description}
The flight data comes from the OpenSky Network, which is a non-profit association that provides open access to real-world air traffic control dataset for research purposes \cite{schafer2014bringing}.
The OpenSky COVID-19 Flight Dataset  \footnote{\href{https://opensky-network.org/community/blog/item/6-opensky-covid-19-flight-dataset}{opensky-network.org}} was made available in April 2020 and is currently updated on a monthly basis, with the purpose of supporting research on the spread of the pandemic and the associated economic impact. This dataset has been used to investigate mobility in the early months of the pandemic \cite{islind2020changes} as well as the pandemic's effect on economic indicators \cite{BankofEngland}. 

The data provides information about the origin and destination airports as well as the date and time of all flights worldwide.
For our analyses we considered domestic flights in the US only. We aggregated the data, to obtain the number of flights between all pairs of airports per day, from the beginning of April until the end of October, 2020. Subsequently, the airports in each state and the number of flights associated with them were combined, to give the number of within and between state flights, on a day to day basis for the whole period.

The number of daily infected cases, which is also used for analysis in this paper, is provided by the open source online repository Opendatasoft \footnote{\href{https://public.opendatasoft.com/explore/dataset/testing-data-covid19-usa/export/?disjunctive.state_name&sort=date}{public.opendatasoft.com/explore/dataset/testing-data-covid19-usa/}}. 

\begin{figure}[tbh!]
\begin{center}
\includegraphics[width=8cm]{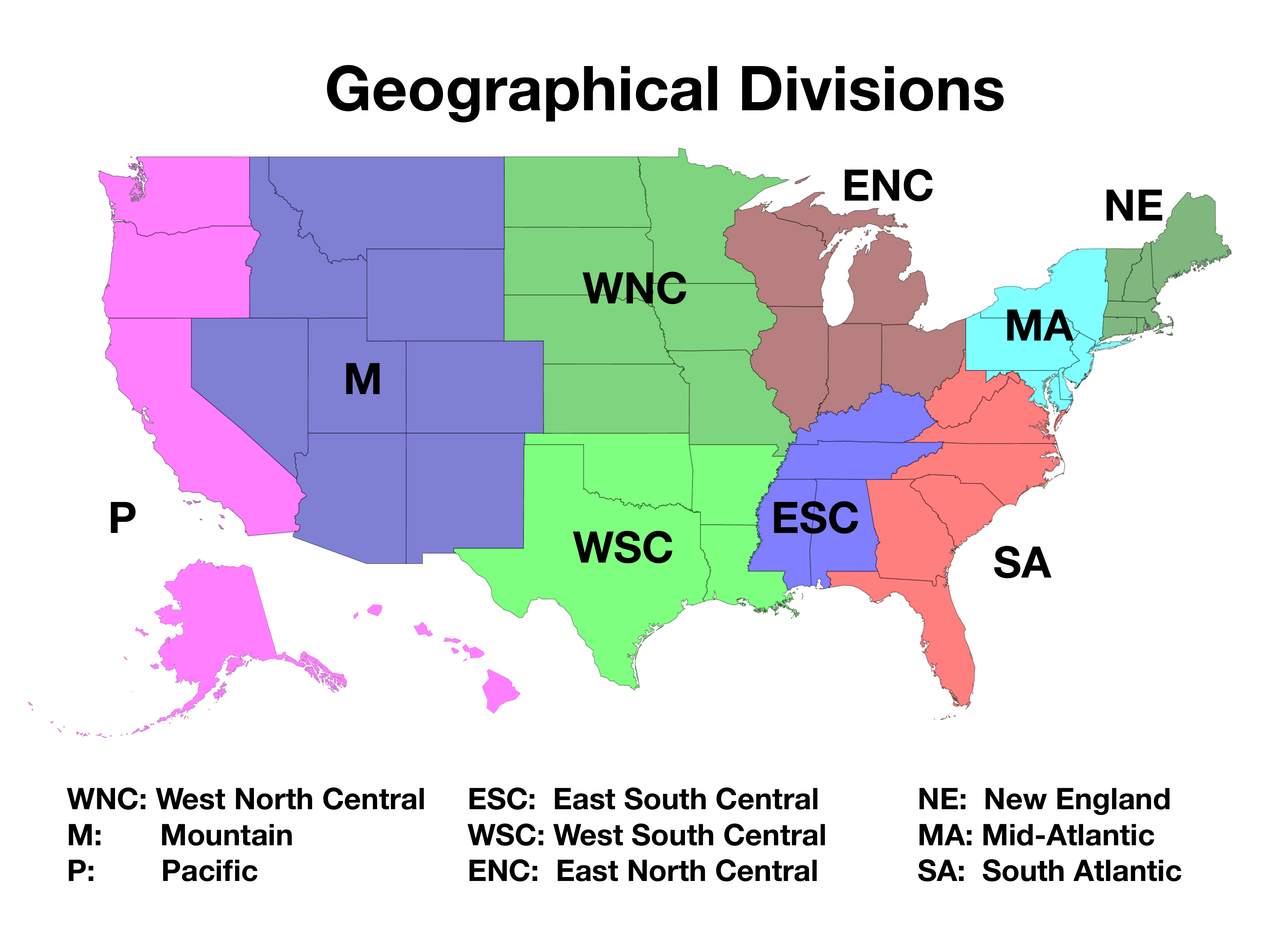}
\end{center}
\caption{Illustration of the geographical divisions of the US used in this study.} \label{fig:Method}
\end{figure}

\subsection{Mathematical modeling}

The states within the US have different population and demographic distribution. A state-by-state mathematical modeling, therefore, is challenged by statistical artifacts. For these reasons we group the states following the census divisions \footnote{\href{https://www2.census.gov/geo/pdfs/reference/GARM/Ch6GARM.pdf}{US Census Bureau}}, as summarized in Table \ref{T-division} and illustrated in Fig.\ref{fig:Method}. 
Note, that contrary to the official definitions, we include Maryland and Delaware in Mid-Atlantic instead of South Atlantic. The main reason is that the population of these two states is more connected to states in Mid-Atlantic, as proven by the diffusion timing of the virus.
    \begin{table*}[tb!]
\begin{center}
\begin{tabular}{ ||p{3.cm}|p{2.2cm}|p{12cm}||}
\hline
\multicolumn{3}{|c|}{Division composition}\\
\hline
Division names & Division code & States within the division\\
\hline
New England & NE & Massachusetts, Connecticut, New Hampshire, Maine, Rhode Island and Vermont \\

Mid-Atlantic & MA & New York, Pennsylvania, New Jersey, \underline{Maryland} and \underline{Delaware} \\

South Atlantic & SA &  Florida, Georgia, North Carolina, Virginia, South Carolina and West Virginia \\

East South Central & ESC & Tennessee, Alabama, Kentucky and Mississippi \\

West South Central & WSC & Texas, Louisiana, Oklahoma and Arkansas \\

East North Central & ENC & Illinois, Ohio, Michigan, Indiana and Wisconsin \\

West North Central & WNC & Missouri, Minnesota, Iowa, Kansas, Nebraska, South Dakota and North Dakota \\

Mountains & M & Arizona, Colorado, Utah, Nevada, New Mexico, Idaho, Montana and Wyoming \\

Pacific & P & California, Washington, Oregon, Hawaii and Alaska \\

 \hline
 \end{tabular}
 \caption{States of the US integrated into 9 divisions. Maryland and Delaware are moved from South Atlantic to Mid-Atlantic.}
 \label{T-division}
 \end{center}
 \end{table*}

Building upon our successful understanding of the COVID-19 temporal evolution  \cite{cacciapaglia2020second} we apply our framework to the US case. Building on that framework we employ the following eRG  set of first order differential equations \cite{Cacciapaglia:2020mjf} to describe the time-evolution of the cumulative number of infected cases within the US divisions: 
\beq
\frac{d \alpha_i}{d t} = \gamma_i \alpha_i \left( 1-\frac{\alpha_i}{a_i} \right) +  \sum_{j\neq i} \frac{k_{ij}}{n_{mi}} (e^{\alpha_j - \alpha_i}  -1)\,,
\label{eq:pandemic}
\eeq
where 
\begin{equation} 
\alpha_i(t) = \rm ln\ \mathcal{I}_i(t) \ ,
\end{equation} 
with $\mathcal{I}_i (t)$ being the cumulative number of infected cases {\it per million} inhabitants for the division $i$ and $\ln$ indicating its natural logarithm. 
These equations embody, within a small number of parameters, the pandemic spreading dynamics across coupled regions of the world via the temporal evolution of $\alpha_i (t)$.  
The parameters $\gamma_i$ and $a_i$ can be extracted by the data within each single wave. The fit methodology is described in \cite{DellaMorte:2020wlc,Cacciapaglia:2020mjf}. 

\begin{figure*}[tbh!]
\begin{center}
\includegraphics[width=5.8cm]{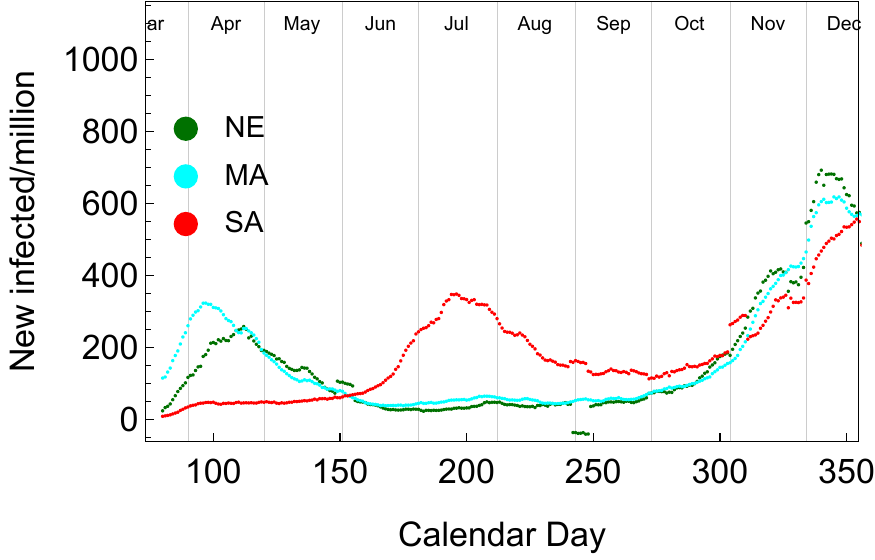}
\includegraphics[width=5.8cm]{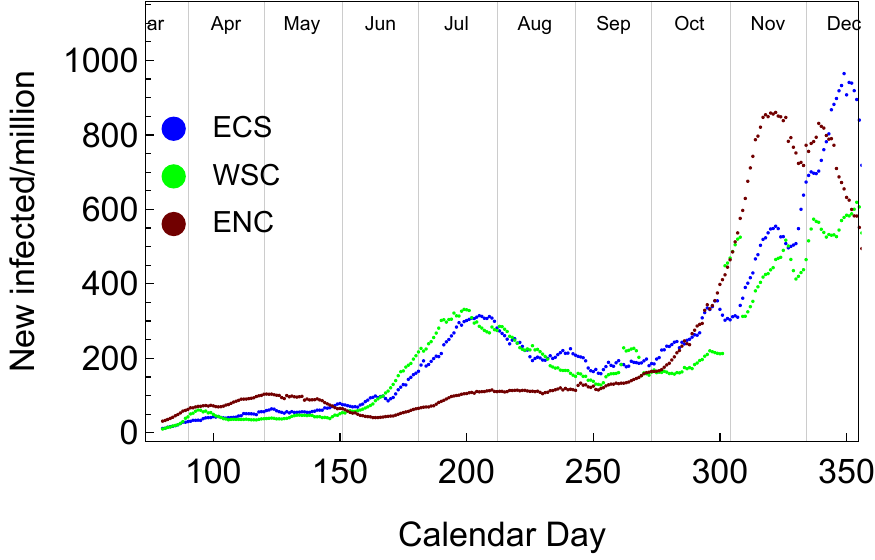}
\includegraphics[width=5.8cm]{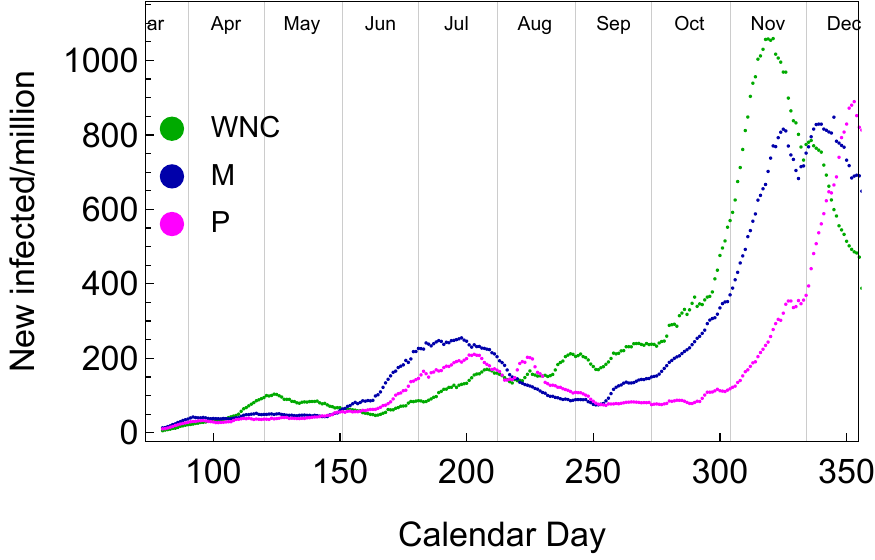}
\end{center}
\caption{Weekly new number of cases for all the 9 divisions. } \label{fig:Newcases}
\end{figure*}

In the US, it is well known that the COVID-19 pandemic started in NE and MA (mainly in New York City) and then spread to the other divisions. Thus, we define the \emph{US first wave} period from March to the end of August as shown in Fig.~\ref{fig:Newcases}. In particular, one observes a peak of new infected in NE and MA around April, while for the other divisions the main peak occurs around July. We also observe an initial feature in the latter divisions that we did not attempt to model except for ENC (mostly located in Chicago) and WNC. For the two latter divisions, we considered these as two independent first wave components. 
The \emph{US second wave} is thus associated with the episode starting in October, 2020.

As a first method, working under the assumption that the US pandemic indeed originated in New York (MA), we first determine the $k_{ij}$ matrix entries between the division MA and the others. The values are chosen to reasonably reproduce the delay between the main peaks of the first wave in pairs of divisions (C.f. the top section in Table~\ref{tab:kappas}). Interestingly, with the exception of NE, the entries of the $k$ matrix are comparable to the ones we used for Europe \cite{cacciapaglia2020second}. For NE, a large coupling is needed due to the tight connections between the two regions, in particular New York City with the neighbouring states and Massachusetts.

As a second method, we used the flight data to estimate the number of travellers between different divisions, under the assumption that the $k_{ij}$ matrix entries are proportional to this set of data. To have a realistic matrix for $k_{ij}$, we first take the mean number of flights from division $i$ to division $j$ during the period from April 1st to May 31st for the first wave, and from September 1st to October 31st for the second wave. Then, we multiply the number of flights by an effective average number of passengers, and normalize it by $10^6$, following the definition of $k_{ij}$ \cite{Cacciapaglia:2020mjf}. For the first wave, the optimal average number of passenger is found to be $10$, while for the second wave we find an optimal value of $5$. Note that these values do not correspond to the actual number of passengers in the flights: in fact, the values of the couplings $k_{ij}$ also take into account the probability of the passengers to carry the infection as compared to the average in the division of origin. When the value is low it might suggest that the sample of passengers in a flight is less infectious than average, as people with symptoms tend not to travel. Controls at airports may also contribute to this. The key information we extract from the flight data is the relative flux of infections among different divisions. 

The results are listed in the middle and bottom sections of Table~\ref{tab:kappas}. We keep the same value from the previous fit only for MA-NE. The reason behind this choice is the tight connection between the two divisions, where most of the human mobility is imputable to road transport.

\begin{table*}[tb!]
\begin{center}
\begin{tabular}{ ||p{2.cm}|p{1.4cm}|p{1.4cm}|p{1.4cm}|p{1.4cm}|p{1.4cm}|p{1.4cm}|p{1.4cm}|p{1.4cm}|p{1.4cm}||}
\hline
\multicolumn{10}{|c|}{$k_{ij}$ values (1st wave fits)}\\
\hline
Division code & NE & MA & SA & ESC & WSC & ENC & WNC & M & P \\
\hline

MA & $ 0.72 $ & $0 $ & $0.0014 $ & $0.00075 $ & $0.0017 $ & $0.0023 $ & $0.0005 $ & $0.002 $ & $0.0053 $ \\
\hline
\multicolumn{10}{|c|}{First wave $k_{ij}$ values (Flight data, from  April 1st to May 31st)}\\
\hline
Division code & NE & MA & SA & ESC & WSC & ENC & WNC & M & P \\
\hline
NE & $ 0 $ & $0.72 $ & $0.0045 $ & $0.00088 $ & $0.00087 $ & $0.0024 $ & $0.00052 $ & $0.00067 $ & $0.00091 $ \\

MA & $ 0.72 $ & $0 $ & $0.019 $ & $0.0056 $ & $0.0041 $ & $0.012 $ & $0.0025 $ & $0.0031 $ & $0.0059 $ \\

SA & $ 0.0043 $ & $0.018 $ & $0 $ & $0.0085 $ & $0.013 $ & $0.019 $ & $0.0057 $ & $0.0050 $ & $0.0067 $ \\

ESC & $ 0.00092 $ & $0.0053 $ & $0.0093 $ & $0 $ & $0.0051 $ & $0.0068 $ & $0.0023 $ & $0.0035 $ & $0.0065 $ \\

WSC & $ 0.00095 $ & $0.0038 $ & $0.014 $ & $0.0055 $ & $0 $ & $0.0092 $ & $0.0054 $ & $0.011 $ & $0.010 $ \\

ENC & $ 0.0025 $ & $0.012 $ & $0.018 $ & $0.0063 $ & $0.0086 $ & $0 $ & $0.0082 $ & $0.0079 $ & $0.0099 $ \\

WNC & $ 0.00038 $ & $0.0022 $ & $0.0056 $ & $0.0019 $ & $0.0046 $ & $0.0070 $ & $0 $ & $0.0055 $ & $0.0027 $ \\

M & $ 0.00050 $ & $0.0020 $ & $0.0042 $ & $0.0026 $ & $0.011 $ & $0.0072 $ & $0.0043 $ & $0 $ & $0.028 $ \\

P & $ 0.00084 $ & $0.0055 $ & $0.0063 $ & $0.0050 $ & $0.010 $ & $0.0092 $ & $0.0033 $ & $0.030 $ & $0 $ \\
\hline
\multicolumn{10}{|c|}{Second wave $k_{ij}$ values (Flight data, from September 1st to October 31st)}\\
\hline
Division code & NE & MA & SA & ESC & WSC & ENC & WNC & M & P  \\
\hline
Region-X & $0.0066$ & $0.028$ & $0.029$ & $0.013$ & $0.019$ & $0.027$ & $0.014$ & $0.03$ & $0.03$ \\ \hline

NE & $ 0 $ & $0.72 $ & $0.0028 $ & $0.00046 $ & $0.00031 $ & $0.0015 $ & $0.00026 $ & $0.00041 $ & $0.00082 $  \\

MA & $ 0.72 $ & $0. $ & $0.011 $ & $0.002 $ & $0.0017 $ & $0.0064 $ & $0.0013 $ & $0.0021 $ & $0.0029 $ \\

SA & $ 0.0026 $ & $0.011 $ & $0 $ & $0.005 $ & $0.005 $ & $0.0096 $ & $0.003 $ & $0.0033 $ & $0.0035 $  \\

ESC & $ 0.00041 $ & $0.0019 $ & $0.0051 $ & $0 $ & $0.0019 $ & $0.0028 $ & $0.00087 $ & $0.0012 $ & $0.0015 $  \\

WSC & $ 0.00028 $ & $0.0015 $ & $0.0049 $ & $0.0018 $ & $0 $ & $0.0028 $ & $0.0016 $ & $0.004 $ & $0.0034 $ \\

ENC & $ 0.0014 $ & $0.0062 $ & $0.0089 $ & $0.0028 $ & $0.003 $ & $0 $ & $0.0039 $ & $0.0043 $ & $0.0045 $  \\

WNC & $ 0.00024 $ & $0.0013 $ & $0.0028 $ & $0.0009 $ & $0.0017 $ & $0.0038 $ & $0 $ & $0.0028 $ & $0.0016 $  \\

M & $ 0.00032 $ & $0.0017 $ & $0.0029 $ & $0.0011 $ & $0.0054 $ & $0.004 $ & $0.0026 $ & $0 $ & $0.014 $  \\

P & $ 0.00074 $ & $0.0028 $ & $0.0032 $ & $0.0014 $ & $0.0046 $ & $0.0041 $ & $0.0018 $ & $0.015 $ & $0 $ \\

 \hline
 \end{tabular}
 \caption{Values of the $k_{ij}$ entries among US divisions. In the top section, the values between Mid-Atlantic (MA) and the other divisions are obtained from fits of the first wave timing. In the central and bottom sections, the complete matrix (except the entries between MA and NE) is obtained using flight data for the first wave (from April 1st to May 31st) and the second (from September 1st to October 31st), respectively. }
 \label{tab:kappas}
 \end{center}
 \end{table*}

By the end of November, we clearly observe a new rise in the number of infections, signalling the onset of a second wave pandemic in the US (see Fig.~\ref{fig:Newcases}). Using our framework, we model and then simulate the second wave across the different US divisions. 
 
\begin{figure}[tbh!]
\begin{center}
\includegraphics[width=7cm]{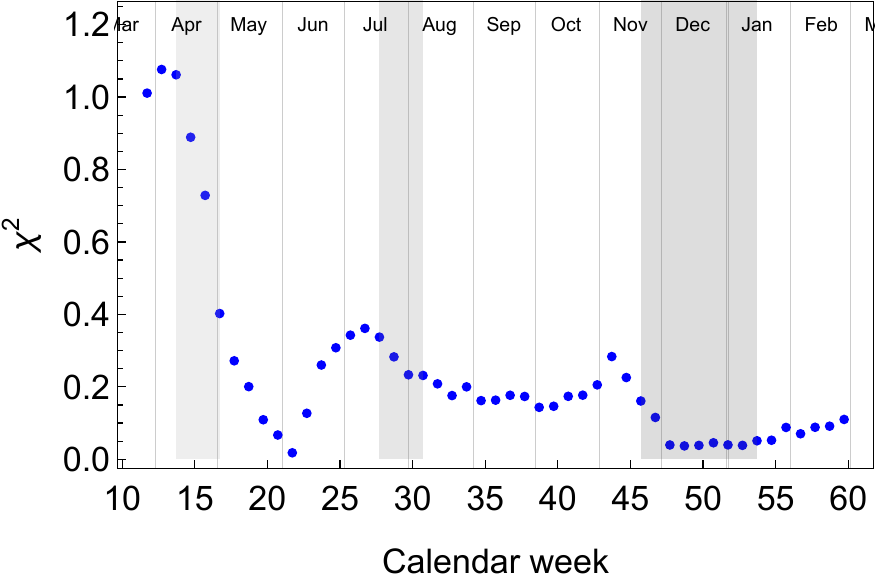}
\end{center}
\caption{Evolution of the uniformity indicator $\chi^2$ over time (weekly basis). The shaded bands indicate the period when epidemic peaks are recorded.} \label{fig:ChiR}
\end{figure}

Finally, to check the geographical diffusion of the virus during the various phases of the pandemic in the US, we define an indicator of the uniformity of the new case incidence \cite{cacciapaglia2020multiwave}. This indicator can be defined week by week via a $\chi^2$-like variable, given by:
\begin{equation}
    \chi^2(t) = \frac{1}{9}\sum^9_{i=1} \left(\frac{\mathcal{I}_i'(t)}{\left<\mathcal{I}'(t)\right>} - 
     1\right)^2\,,
\end{equation}
where $\mathcal{I}_i'(t)$ is the number of new cases per week in division $i$ at time $t$ and $\left<\mathcal{I}'(t)\right>$ the mean of the same quantity in the 9 divisions.
The parameter $\chi^2$ quantifies the geographical diffusion of the SARS-CoV-2 virus in the US: the smaller its value, the more uniform the pandemic spread within the whole country.
The result is shown in Fig. \ref{fig:ChiR}: during the first peak in April (light gray shade), the value of $\chi^2$ is large, signalling that the epidemic diffusion is localized in a few divisions; during the second peak of the first wave (gray shade), the value has dropped, signalling that the epidemic has been spreading to all divisions. Finally, the data for the ongoing second wave (dark gray shade) shows that $\chi^2$ is dropping towards zero, as expected for a more  diffuse incidence of infections.

\subsection{Vaccine deployment and implementation}

Various vaccines have been developed for the COVID-19 pandemic, and their deployment in the US has already started on December 14th \footnote{\href{https://www.washingtonpost.com/nation/2020/12/14/first-covid-vaccines-new-york/}{https://www.washingtonpost.com}}. The effect of the immunization due to the vaccine has been studied in the context of compartmental models, like SEIR \cite{vaccine}. In our mathematical model, the simplest and most intuitive effect is a reduction of both the total number of infections during a single wave, $a_i$, and/or the effective diffusion rate of the virus $\gamma_i$, in each division.

\begin{figure}[htb!]
    \centering
    \includegraphics[scale=0.8]{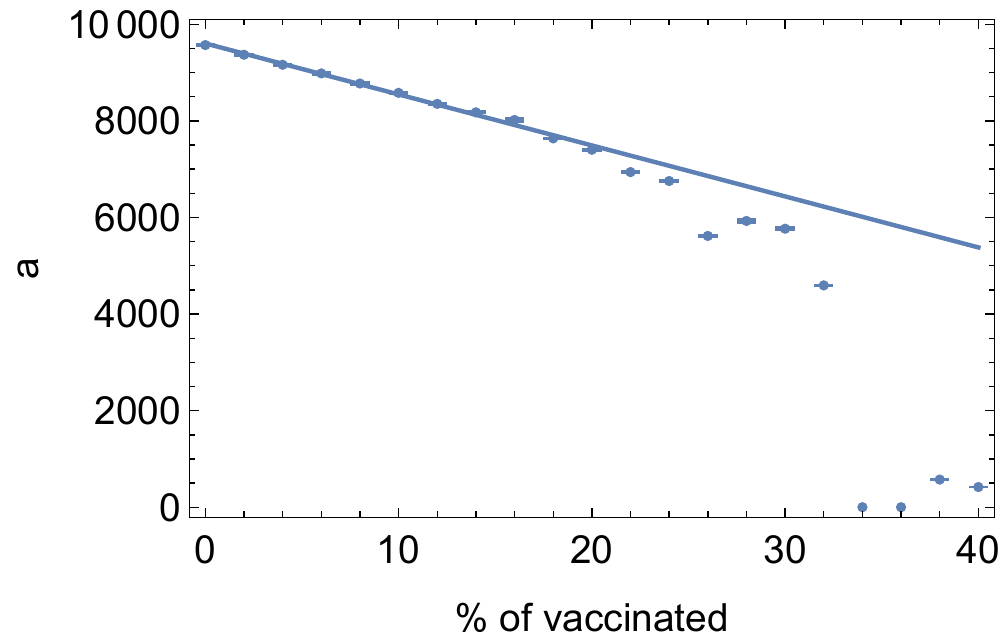}
    \includegraphics[scale=0.8]{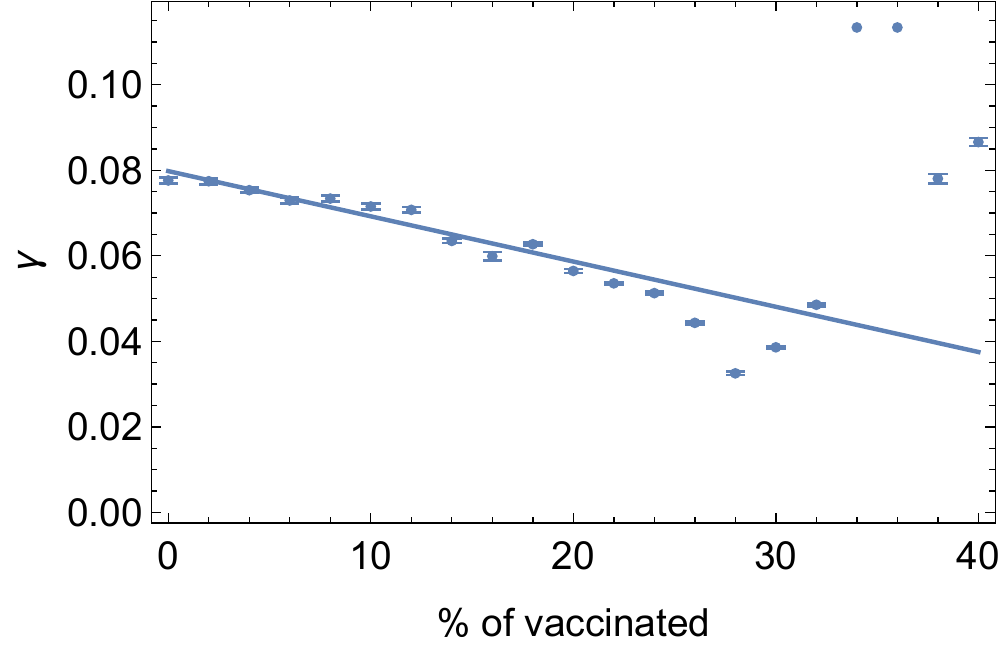}
    \caption{$a$ and $\gamma$ fit parameters versus initial percentage of vaccinated nodes for $\gamma_\ast= 0.6$ and $\epsilon_\ast = 0.4$.} 
    \label{fig:vaxsim}
\end{figure}

To validate this working hypothesis, and understand how the vaccination of a portion of the population affects the values of $a$ and $\gamma$ in the eRG framework, we studied the effect of immunization in a simple percolation model, which has been shown to be in the same class of universality as simple compartmental models \cite{Cardy_1985}. 
To do so, we set up a Monte-Carlo simulation, consisting of a square grid whose nodes are associated to a susceptible individual. Each node can be in four exclusive states: Susceptible (S), Infected (I), Recovered (R) or Vaccinated (V). At each step in time in the simulation, for each node we generate a random number $r$ between 0 and 1: if the node is in state S in proximity to a node in state I and $r<\gamma_\ast$, we switch its state to I, else it remains S; if the node is in state I and $r<\epsilon_\ast$, we switch its state to R, else it remains I; if the node is in state R or V, it will not change.
This model reproduces the diffusion of the infection, where $\gamma_\ast$ is the infection probability on the lattice and $\epsilon_\ast$ is the recovery rate. Finally, we fit the data from the simulation to the solution of a simple eRG equation to extract $\gamma$ and $a$. The vaccination is implemented by setting a random fraction $R_v$ of nodes to the state V before the simulation starts. The values of $a$ and $\gamma$ as a function of the fraction of vaccinations are shown in Fig.\ref{fig:vaxsim}: we observe that both parameters are reduced by the same percentage as the vaccination up to $R_v \lesssim 25$\%. Above this value of vaccinated nodes, the simulation is unstable and the result cannot be trusted. This result, nevertheless, demonstrates that the vaccination reduces both parameters $a$ and $\gamma$ proportionally reinforcing our expectation.

In a realistic scenario, the vaccination of the population can only be implemented in a gradual way, so that the total vaccination campaign has a duration in time. 
We can thereby assume that a fraction $R_v$ of the population is vaccinated in a time interval $\Delta_t$. The rate of vaccinations is therefore $c = R_v/\Delta_t$. This implies that the variation in $\gamma$, during the time interval from $t_v$ to $t_v + \Delta_t$, is given by:
\begin{equation}
\frac{d \gamma (t)}{dt} = - c\ \gamma(t_v)\,,
\end{equation}
where $\gamma (t_v)$ is the effective infection rate before the start of the vaccination campaign.
The solution for the time-dependent effective infection rate is
\begin{eqnarray}
\gamma(t) &=& \gamma (t_v) [1 - c (t-t_v)]\,,
\end{eqnarray}
until $t=t_v + \Delta_t$, after which $\gamma$ remains constant again at a reduced value $\gamma (t_v)\ (1-R_v)$.

To find the variation of $a(t)$ within the vaccination interval $t_v$ to $t_v + \Delta_t$, we assume that the not-yet-infected individuals are vaccinated at the same rate $c$ as the total population. Thus, at any given time, the variation in the number of individuals that will be exposed to the infection, $e^{a(t)}$, is proportional to the difference $e^{a(t)} - e^{\alpha(t)}$. This leads to the following differential equation:
\begin{equation}
\frac{d a(t)}{dt} = - c\  (1 - e^{\alpha(t)-a(t)})\,. 
\end{equation}
This equation needs to be solved in a coupled system with the eRG one.
Note that the derivative is zero outside of the time interval $[t_v, t_v + \Delta_t]$.
In the numerical solutions for the effect of the vaccine, we will add one equation for each $a_i (t)$, assuming that the vaccination rate $c$ is the same in all divisions.

\begin{center}
\begin{table*}[tb!]
\begin{tabular}{ ||p{3cm}|p{.8cm}||p{3.2cm}|p{3.2cm}||p{2.8cm}|p{2.8cm}||}
\hline
\multicolumn{2}{||c||}{} & \multicolumn{2}{c||}{First wave parameters (fitted)} & \multicolumn{2}{c||}{Second wave parameters }\\
\hline
Division & \multicolumn{1}{c||}{Code} &
 \multicolumn{1}{c|}{$a$} & \multicolumn{1}{c||}{$\gamma$} &  \multicolumn{1}{c|}{$a$} & \multicolumn{1}{c||}{$\gamma$} \\
\hline

New England & NE & $9.397(7)$ & $0.416(7)$ &$11.006$&$0.214$\\

Mid-Atlantic & MA & $9.496(7)$ & $0.516(9)$ &$10.882$&$0.206$ \\

South Atlantic & SA & $9.691(5)$ & $0.370(3)$ &$10.885$&$0.185$\\

East South Central & ESC & $9.63(2)$ & $0.331(6)$ &$11.201$&$0.207$\\

West South Central & WSC & $9.720(6)$ & $0.340(3)$ &$10.713$&$0.213$\\

East North Central & ENC & $8.88(3)$ & $0.300(8)$ &$11.074$&$0.250$\\

West North Central & WNC & $8.66(2)$ & $0.342(6)$ &$11.060$&$0.263$\\

Mountain & M & $9.478(9)$ & $0.330(4)$ &$11.089$&$0.213$\\

Pacific & P & $9.33(1)$ & $0.291(5)$ &$11.535$&$0.171$\\
 \hline
 \end{tabular}
 \caption{Parameters of the eRG model for the first and second wave in the 9 divisions. For the first wave, we report the values from the fit, including the $1\sigma$ error. For the second wave, the values are chosen to reproduce the current data, adjourned to December 16th.}
 \label{tab:params}
 \end{table*}
 \end{center}

\section{Results} 
 
\subsection{Validating the eRG on the first wave data}

The epidemic data (C.f. Fig.\ref{fig:Newcases}) shows that the MA division (New York City) was first hit hard by the COVID-19 pandemic, and was followed closely by NE. The other divisions witnessed a comparable peak of new infections 3-4 months later. Note that we are using cases normalized per million to facilitate the comparison between divisions with varying population. As a first study, we want to test the $eRG$ equations \eqref{eq:pandemic} against the hypothesis that the epidemic has been diffusing from MA to the other divisions. The parameters $a_i$ and $\gamma_i$ are fixed by fitting the data, as shown in Table \ref{tab:params}. Thus, the timing of the peaks in the divisions is determined by the entries of the $k_{ij}$ matrix. Determining all 81 entries from the data is not possible, as we only have 9 epidemiological curves. Thus, we assume that only the couplings between the source MA and any other division are responsible. The results of the fits are shown in the top block of Table \ref{tab:kappas}, and will be used as a control benchmark. 

Except for $k_{21}$ that links NE and MA, all the other $k_{2j}$ are of order $10^{-3}$, thus confirming the range we found for the European second wave \cite{cacciapaglia2020second}. The value of $k_{21}$ is of order unity, which implies that there is a stronger connection between the two divisions. This may be explained by the fact that there exist a significant flow of people between New York City and the neighbouring states (including Massachusetts) in New England. Work commutes and weekend travelling by car explains the required high number of travellers per week.
Another interesting feature is the presence of a small peak of infections for ENC and WNC, around March. 
This feature cannot originate from the MA division, as that would imply a $k$-value of order $10$, which is clearly unrealistic \cite{Cacciapaglia:2020mjf}. The only viable solution is that the epidemic hit these two divisions from abroad. On the other hand, the second peak observed around August can be explained by the interaction with MA.

The values of $k_{ij}$ are, in principle, determined by the flow of people between different divisions. Thus, we could use any set of mobility data \cite{Yang_2020} to estimate the relative numbers of the entries, while the normalization also depends on the effective infection power of the traveling individuals and it can be determined from the data. With the help of mobility data, we can reduce the 81 parameters to a single one. Due to the large distances across divisions, we decided to focus on the flight data, as described in the methodology section.  
The values of the entries are reported in the middle section of Table \ref{tab:kappas}. 
Note that for MA-NE we used the same value obtained from the previous fit, as the people's flow is mostly dominated by land movements.

\begin{figure*}[tbh!]
\begin{center} \begin{tabular}{cc}
\phantom{xxxx} \mbox{\bf \large First wave} & \phantom{xxx} \mbox{\bf \large Second wave} \\
\includegraphics[width=8cm]{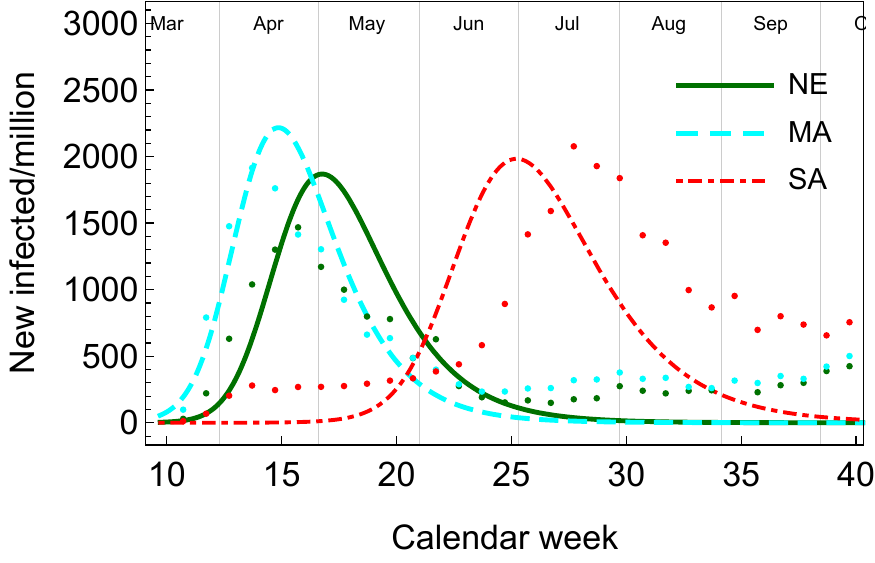} & \includegraphics[width=8cm]{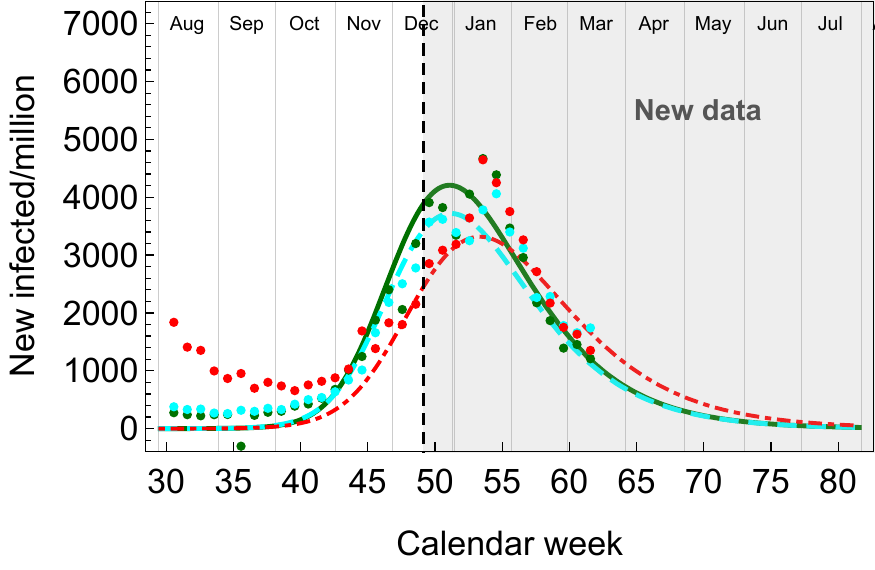}\\
\includegraphics[width=8cm]{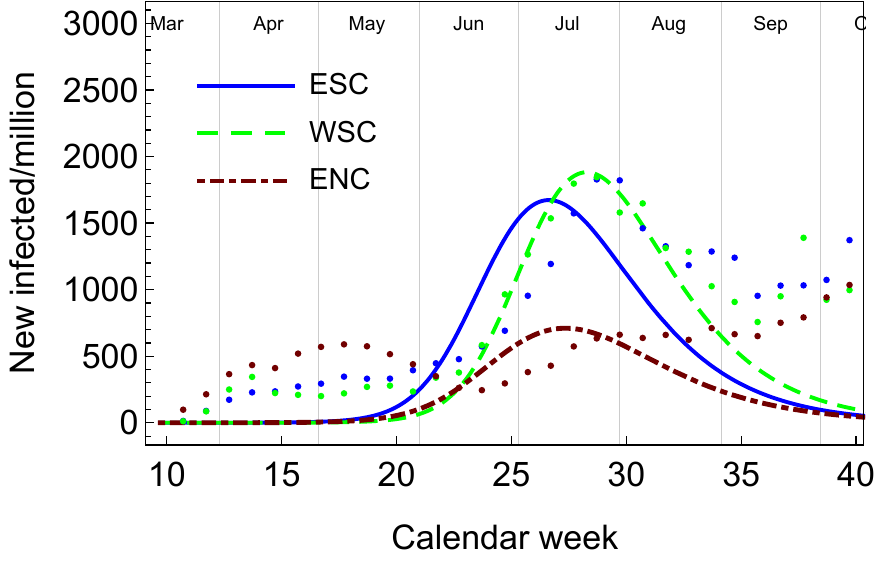} & \includegraphics[width=8cm]{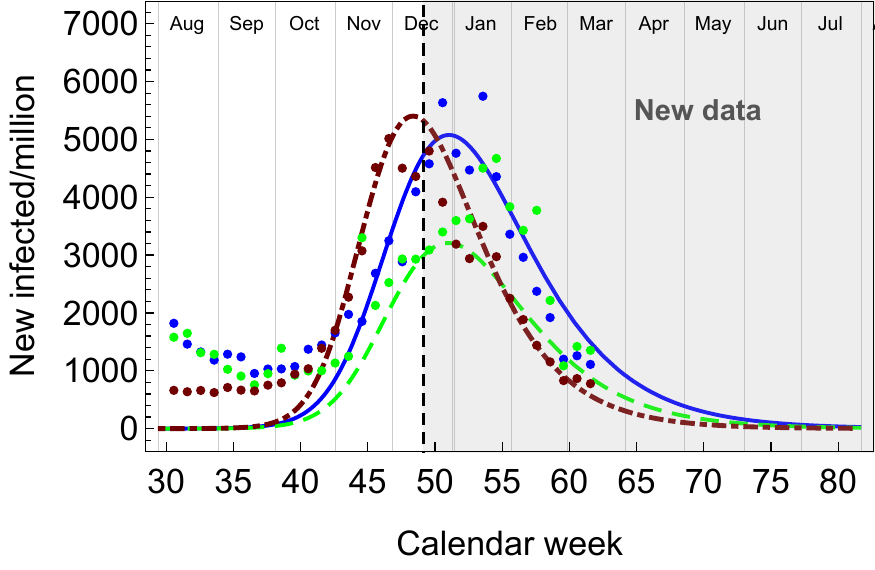}\\
\includegraphics[width=8cm]{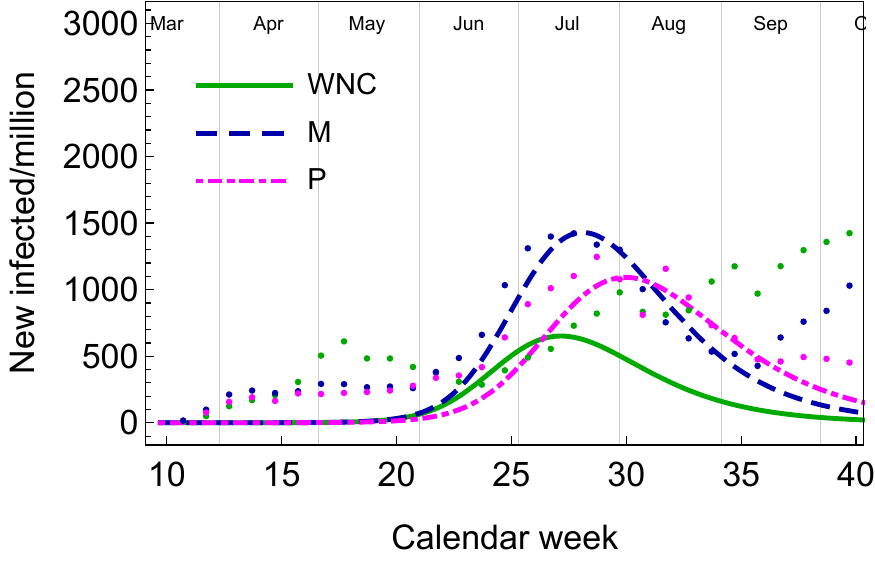} & \includegraphics[width=8cm]{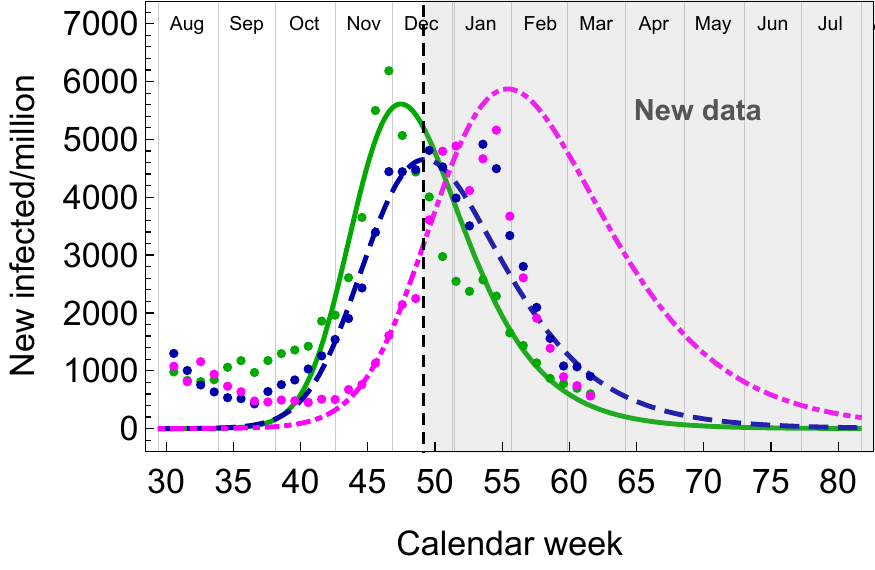}
\end{tabular}\end{center}
\caption{Simulation of the spread of the first wave (left plots) and the second wave (right plots) using flight-data-derived kappa matrix. For the first wave, MA is used as a seed region, while for the second wave a combination of the first waves among divisions acts as the seed region (Region-X). The vertical dashed lines in the right column plots mark the date when the simulation was done. The data points in the grayed region where not used to tune the eRG solutions.} \label{fig:SimFlights}
\end{figure*}

Using this matrix of $k_{ij}$ to simulate the spread of the first wave across the country, as originating from MA, we obtain the curves in the left panels of Fig.\ref{fig:SimFlights}. For nearly all divisions, we obtain the correct timing for the peak, with the exception of SA and ESC (for ENC and WNC, the anomaly may be linked to the presence of a mild early peak and the absence of a prominent second peak).
The results are more accurate for divisions far from MA, thus validating the method as the diffusion of the virus seems to depend on the people travelling (by air) among divisions. For SA, the predicted curve is substantially anticipated compared to the data: this discrepancy may be explained by the presence of an air hub in Atlanta, GE, so that many of the passengers of flights landing there do not stop in the division but instead take an immediate connecting flight.

\subsection{Understanding the second wave}

The US states are currently witnessing a second wave, which is ravaging in all the 9 divisions with comparable intensity.
Previous studies in the eRG framework have uncovered two possible origins for an epidemic wave to start: one is the coupling with an external region with a raging epidemic \cite{Cacciapaglia:2020mjf}, the second is the instability represented by a strolling phase in between waves \cite{cacciapaglia2020evidence,cacciapaglia2020multiwave}. 
We have shown that the former mechanism can account for the peak structure during the first wave.

As a first step, we will try to use the same method to understand the second wave.
 Since travelling to the US from abroad has been strongly reduced and regulated, we will consider the divisions that witnessed a peak in July-August as source for the second wave.
To this purpose, we define a Region-X \cite{Cacciapaglia:2020mjf} as an average sum of all the divisions with a pandemic peak occurring in the July-August period. The parameters are chosen to reproduce the number of cases in the totality of the relevant 7 divisions (SA, ESC, NSC, ENC, WNC, M and P) normalized by the total population. For each division, we optimized $a_i$ and $\gamma_i$ to reproduce the current data adjourned at December 16th (C.f. Table \ref{tab:params}). For the couplings $k_{ij}$ we use the flight data, except for the usual MA-NE couplings (C.f. bottom section of Table \ref{tab:kappas}). 
Finally, the $k_{0j}$ connecting the 9 divisions to the source Region-X are computed by summing the $k$ entries between the division $j$ and the 7 divisions used to model Region-X (also derived from flight data). 

The results of the eRG equations are shown in the right panels of Fig. \ref{fig:SimFlights}, showing a good agreement. The values of the $k_{0j}$ of the Region-X are one order of magnitude larger than the others. This fact can be interpreted by the presence of hotspots in each division which also contribute significantly to the new wave. In other words, traveling among divisions cannot be the only responsible factor for the onset of the second wave in the US. This hypothesis can also be validated by studying the uniformity of the distribution of the new infections in various states during the three peaks, as shown in Fig.\ref{fig:ChiR}. Comparing the three peak regions, we see that the uniformity indicator is systematically decreasing, thus indicating a more geographically uniform presence of the virus.

It is also interesting to notice that the value of $\gamma_i$ for the second wave is systematically smaller than the infection rate during the first wave. This is in agreement with the results we found in \cite{cacciapaglia2020evidence,cacciapaglia2020multiwave}, where we modelled the multi-wave structure of the pandemic via an instability inside each region. The result of this simple analysis supports the hypothesis that the virus is now endemic for all states in the US, thus a multi-wave pattern will continue to emerge. Traveling among states (or divisions) is less relevant at this stage.

The result of our eRG analysis shows that the current wave will end in March-April 2021. Note, however, that we have not taken into account the potential disastrous effect of the Christmas and New Year holidays, which could lead to an increase in the infection rates. In some divisions there is a increase at the end of November, which can be attributed to the Thanksgiving holiday.

\subsection{Effect of the current vaccination strategy}

\begin{figure*}[tb!]
    \centering
\begin{tabular}{cc}
\phantom{xxxx} \mbox{\bf \large South Atlantic} & \phantom{xxxx}\mbox{\bf \large West North Central} \\
    \includegraphics[scale=0.8]{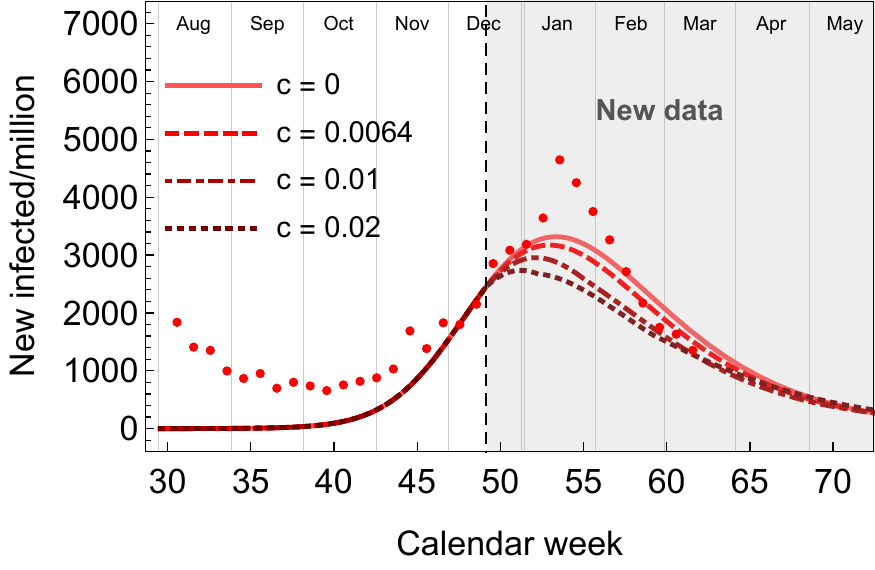} &
    \includegraphics[scale=0.8]{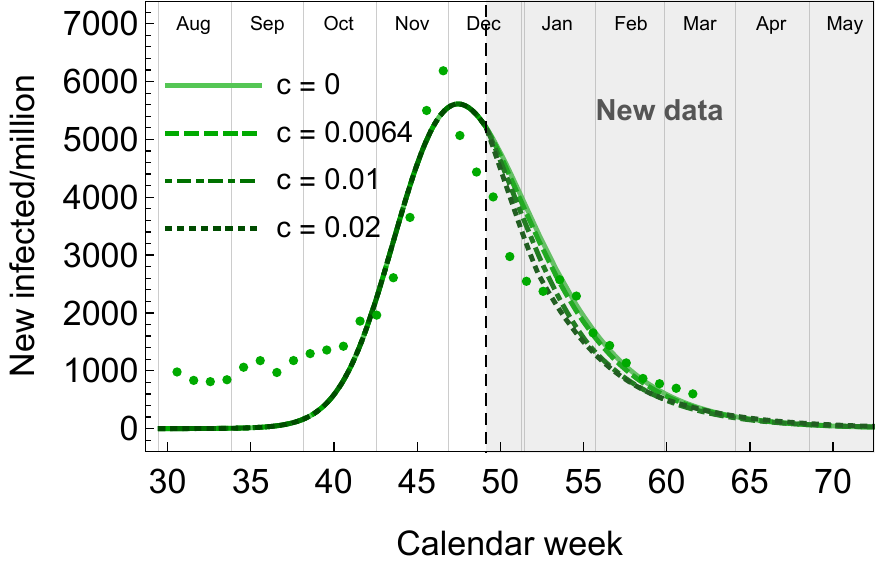}
\end{tabular}    
    \caption{Evolution of the number of infections without vaccination ($c = 0$) and with a vaccination rate of $0.64\%$/week, $1\%$/week and $2\%$/week starting on December 14th and stopping at $20\%$ of the population vaccinated. We show the results for two sample divisions: South Atlantic and West North Central.} 
    \label{fig:vax}
\end{figure*}

Following the development of multiple vaccines for the SARS-COV-2 virus \footnote{\href{https://www.nature.com/articles/d41586-020-03626-1}{https://www.nature.com/articles/d41586-020-03626-1}}, vaccination campaigns have started in many countries, including the US. This will influence the development of the current wave, and help in curbing the future ones.
The vaccination campaign started on December 14th in the US  \footnote{\href{https://www.washingtonpost.com/nation/2020/12/14/first-covid-vaccines-new-york/}{https://www.washingtonpost.com}}. We also know that the US has purchased 100 million doses from Pfizer (plus an additional 100 million from Moderna) \footnote{\href{https://www.forbes.com/sites/leahrosenbaum/2020/12/07/the-us-has-ordered-enough-of-pfizers-covid-19-vaccine-for-50-million-people-but-thats-all/}{https://www.forbes.com}}, so that at least $20\%$ of the population may be vaccinated in this first campaign. 
As of December 28, $0.64\%$ of the total population has been vaccinated \footnote{\href{https://ourworldindata.org/covid-vaccinations}{https://ourworldindata.org}} in a little over one week, thus in our study we will use this as a benchmark weekly rate.
The data listed above defines our starting benchmark for the current vaccination campaign.

To study the effect of the vaccinations, we have solved the eRG equations for the second wave, with the addition of the reduction of $a_i$ and $\gamma_i$, as detailed in the methodology section. We show the result for two sample divisions in Fig.\ref{fig:vax} (dashed curves) as compared to the same solutions without vaccines (solid curves). A vaccination at a $0.64\%$ rate per week does not affect the peak of new infections.
As a reference, we also increased the vaccination rates to $1\%$ and $2\%$: in these cases, an important flattening of the epidemic curve can be observed for SA, where the vaccination started early compared to the peak of infections. This situation may be realized, as the vaccine is being administered to the population that is more at risk of being infected by the virus. In the other extreme case, represented by WNC, the vaccine is ineffective in changing the current wave because the peak has already been attained before the vaccination campaign started. 

Our results confirm that the current vaccination strategy, which is performed during a peak episode, is not effective to substantially slow down the spread of the virus. On the other hand, the effectiveness for future waves is not a question. It would be, in fact, very efficient to be able to administer the vaccine to a larger portion of the population before the start of the next wave.

\subsection{Update of the vaccination to the first quarter 2021}

As shown in the right column in Fig.~\ref{fig:SimFlights}, our simulation of the second wave, done in mid December 2020, reproduces very well the epidemiological data up to March 17, 2021. The only exception is Pacific, which has seen a sharper drop in the number of new infections. Furthermore, one can clearly see a rebounce in January that can be accredited to the Christmas holidays. Nevertheless, this small effect does not have a significant impact on the agreement of our prediction with the data.

In the first quarter of 2021, the vaccination campaign has also taken off steadily, with nearly a quarter of the US population having received at least one shot of vaccine. Furthermore, since February 27 the FDA has authorised the use of the Janssen mono-dose vaccine~\footnote{https://www.fda.gov/emergency-preparedness-and-response/coronavirus-disease-2019-covid-19/janssen-covid-19-vaccine}, which is now being administered together with the two-dose Pfizer and Moderna vaccines. The data show that the rate of vaccinations has been increasing approximately linearly with time, thus we updated the prediction to take into account a vaccination fraction $c(t)$ growing linearly with time:
\begin{equation}
    c(t) = u \; t\,,
\end{equation}
where the numerical value for each division are shown in Table~\ref{tab:Vaxparams}.

\begin{figure*}[tbh!]
\begin{center} 
\includegraphics[scale=0.54]{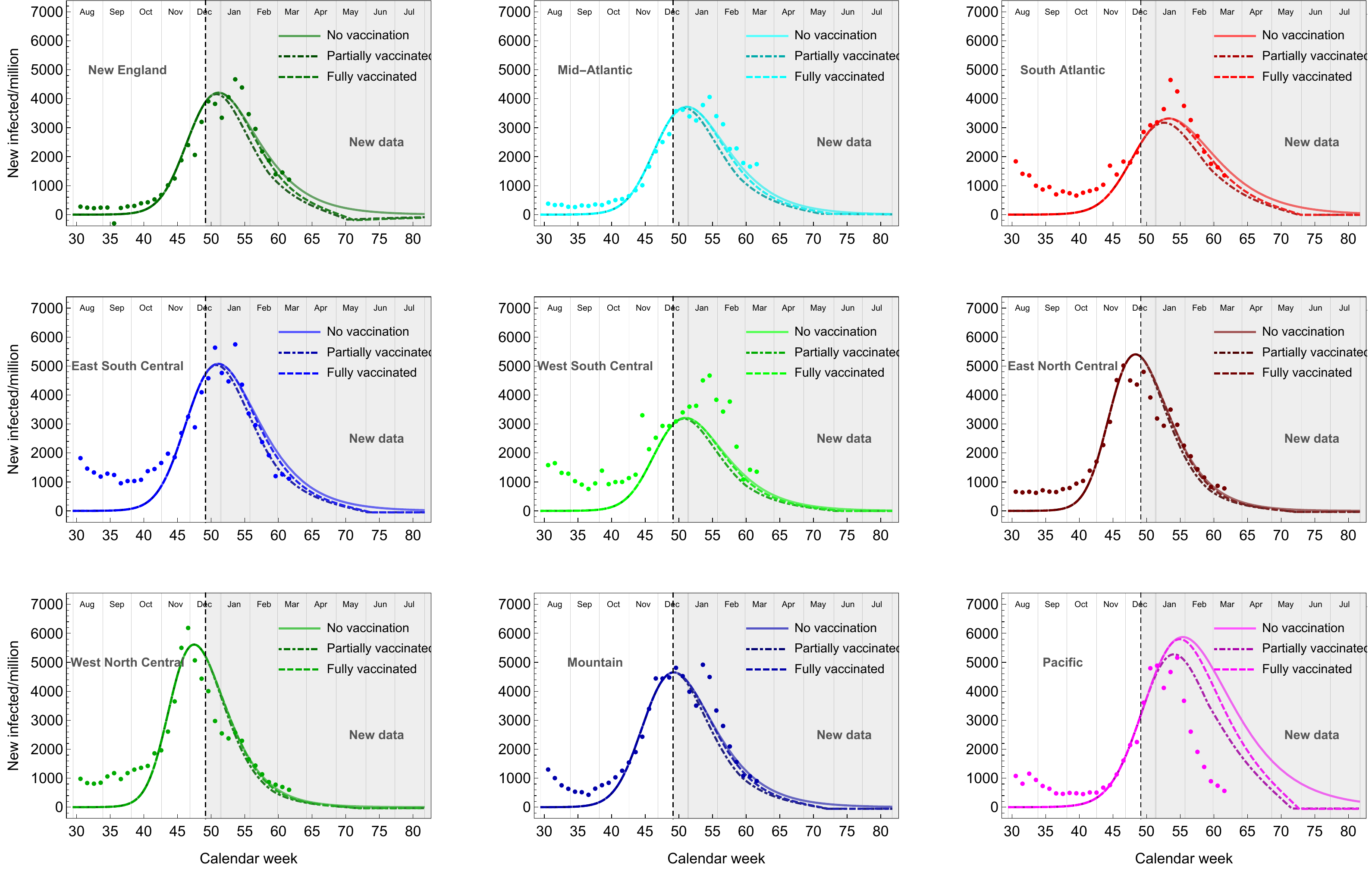}
\end{center}
\caption{Results of the eRG solutions for the second wave with a vaccination campaign based on the data. Here, we consider a vaccination rate linearly increasing in time, with slopes given in Table \ref{tab:Vaxparams}. The eRG parameters are the same used for Fig.~\ref{fig:SimFlights}, based on data until December 28, 2020.} \label{fig:VaxDivisions}
\end{figure*}

The new results are shown in Fig.~\ref{fig:VaxDivisions} for the 9 divisions. We consider both people that received at least one shot (partial vaccination, in dash-dotted lines) and fully vaccinated ones (dashed lines), with an offset of 4 weeks between the beginning of the two. We consider them as two extreme cases, defining a systematic error in our account of vaccinations. In all divisions, the effect is minor, as the vaccination campaign has started too close to the peak of the second wave. The only exception is Pacific, where taking into account vaccinations substantially improves the agreement with the data.

The updated results confirm that a vaccination campaign operating during a wave will not significantly affect the timing and height of the peak. Social distancing and containment measures remain necessary. Conversely, vaccinating a large portion of the population will certainly curb the eventual next wave.

\begin{center}
\begin{table*}[tb!]
\begin{tabular}{ ||p{3cm}|p{.8cm}||p{3.2cm}|p{3.2cm}||p{3.2cm}|p{3.2cm}||}
\hline
\multicolumn{2}{||c||}{} & \multicolumn{2}{c||}{Percentage of the population vaccinated }& \multicolumn{2}{c||}{Vaccination rate slope $u$}\\
\hline
Division & \multicolumn{1}{c||}{Code} &
 \multicolumn{1}{c|}{Partial vaccination} & \multicolumn{1}{c||}{Full vaccination}&
 \multicolumn{1}{c|}{Partial vaccination} & \multicolumn{1}{c||}{Full vaccination}\\
\hline

New England & NE & $30.1$ & $16.4$ & $0.00450$ & $0.00641$\\

Mid-Atlantic & MA & $27.8$ & $14.1$ & $0.00415$ & $0.00554$\\

South Atlantic & SA & $24.3$ & $13.5$ & $0.00363$ & $0.00530$\\

East South Central & ESC & $23.1$ & $12.6$ & $0.00345$ & $0.00492$\\

West South Central & WSC & $23.0$ & $11.9$ & $0.00344$ & $ 0.00467$\\

East North Central & ENC & $25.6$ & $14.7$ & $0.00382$ & $0.00575$\\

West North Central & WNC & $26.3$ & $14.9$ & $0.00393$ & $0.00584$\\

Mountain & M & $25.7$ & $14.8$ & $0.00386$ & $0.00579$\\

Pacific & P & $ 26.6$ & $14.0$ & $0.00397$ & $0.00548$\\

 \hline
 \end{tabular}
 \caption{Percentage of the population vaccinated with at least one dose and with two doses in each division as of the date of 24th of march 2021.}
 \label{tab:Vaxparams}
 \end{table*}
 \end{center}

\section{Discussion}
\label{sec:5}

In this paper we employ the epidemic Renormalization Group (eRG) framework in order to understand, reproduce and predict the diffusion of the COVID-19 pandemic across the US as well as the effect of vaccination strategies. By using flight data, we are able to see the changes in mobility across the divisions, and observe how these changes affect the spread of the virus. Furthermore, we show that the impact of the vaccination campaign on the current wave of the pandemic in the US is marginal. Based on that, the importance of social distancing is still relevant. Furthermore, we demonstrate that the current wave is due to the endemic diffusion of the virus. Therefore, building upon our previous results \cite{cacciapaglia2020multiwave}, in order to control the next pandemic wave the number of daily new cases per million must be around or less than 10-20 during the next inter-wave period. This conclusion is further corroborated in \cite{priesemann2020calling} for Europe.

We learnt that the number of infected individuals in the current wave are not affected measurably by the vaccination campaign. However, it is foreseeable that it will impact specific compartments such as the overall number of deceased individuals. Our study included an immunization rate between 0.64\% to 2\% of the total population each week. 
We also updated the results with the actual rates of vaccination in the different divisions, as of March 24, 2020. The results of our eRG model agree remarkably well with the new data from December 28, 2020, to March 17, 2021.
To curb the current and the next waves, our results indisputably show that vaccinations alone are not enough and strict social distancing measures are required until sufficient immunity is achieved.  

Our results should be employed by governments and decision makers to implement local and global measures and, most importantly, the results of this paper can be used as a foundation for vaccination campaign strategies for governments.

Given that pandemics are recurrent events, our results go beyond COVID-19 and are universally applicable. What we have seen in the data for the US is that it started in New York and, from there, it diffused to the rest of the country. It is, therefore, important to contain future pandemics at an early stage.

\bibliography{biblio}

\begin{thebibliography}{37}%
\makeatletter
\providecommand \@ifxundefined [1]{%
 \@ifx{#1\undefined}
}%
\providecommand \@ifnum [1]{%
 \ifnum #1\expandafter \@firstoftwo
 \else \expandafter \@secondoftwo
 \fi
}%
\providecommand \@ifx [1]{%
 \ifx #1\expandafter \@firstoftwo
 \else \expandafter \@secondoftwo
 \fi
}%
\providecommand \natexlab [1]{#1}%
\providecommand \enquote  [1]{``#1''}%
\providecommand \bibnamefont  [1]{#1}%
\providecommand \bibfnamefont [1]{#1}%
\providecommand \citenamefont [1]{#1}%
\providecommand \href@noop [0]{\@secondoftwo}%
\providecommand \href [0]{\begingroup \@sanitize@url \@href}%
\providecommand \@href[1]{\@@startlink{#1}\@@href}%
\providecommand \@@href[1]{\endgroup#1\@@endlink}%
\providecommand \@sanitize@url [0]{\catcode `\\12\catcode `\$12\catcode
  `\&12\catcode `\#12\catcode `\^12\catcode `\_12\catcode `\%12\relax}%
\providecommand \@@startlink[1]{}%
\providecommand \@@endlink[0]{}%
\providecommand \url  [0]{\begingroup\@sanitize@url \@url }%
\providecommand \@url [1]{\endgroup\@href {#1}{\urlprefix }}%
\providecommand \urlprefix  [0]{URL }%
\providecommand \Eprint [0]{\href }%
\providecommand \doibase [0]{http://dx.doi.org/}%
\providecommand \selectlanguage [0]{\@gobble}%
\providecommand \bibinfo  [0]{\@secondoftwo}%
\providecommand \bibfield  [0]{\@secondoftwo}%
\providecommand \translation [1]{[#1]}%
\providecommand \BibitemOpen [0]{}%
\providecommand \bibitemStop [0]{}%
\providecommand \bibitemNoStop [0]{.\EOS\space}%
\providecommand \EOS [0]{\spacefactor3000\relax}%
\providecommand \BibitemShut  [1]{\csname bibitem#1\endcsname}%
\let\auto@bib@innerbib\@empty
\bibitem [{\citenamefont {Perc}\ \emph {et~al.}(2020)\citenamefont {Perc},
  \citenamefont {Gori\v{s}ek~Miksi\'{c}}, \citenamefont {Slavinec},\ and\
  \citenamefont {Sto\v{z}er}}]{Perc2020}%
  \BibitemOpen
  \bibfield  {author} {\bibinfo {author} {\bibfnamefont {M.}~\bibnamefont
  {Perc}}, \bibinfo {author} {\bibfnamefont {N.}~\bibnamefont
  {Gori\v{s}ek~Miksi\'{c}}}, \bibinfo {author} {\bibfnamefont {M.}~\bibnamefont
  {Slavinec}}, \ and\ \bibinfo {author} {\bibfnamefont {A.}~\bibnamefont
  {Sto\v{z}er}},\ }\href {\doibase https://doi.org/10.3389/fphy.2020.00127}
  {\bibfield  {journal} {\bibinfo  {journal} {Frontiers in Physics}\ }\textbf
  {\bibinfo {volume} {8}},\ \bibinfo {pages} {127} (\bibinfo {year}
  {2020})}\BibitemShut {NoStop}%
\bibitem [{\citenamefont {H\^{a}ncean}\ \emph {et~al.}(2020)\citenamefont
  {H\^{a}ncean}, \citenamefont {Perc},\ and\ \citenamefont
  {Juergen}}]{Hancean2020}%
  \BibitemOpen
  \bibfield  {author} {\bibinfo {author} {\bibfnamefont {M.-G.}\ \bibnamefont
  {H\^{a}ncean}}, \bibinfo {author} {\bibfnamefont {M.}~\bibnamefont {Perc}}, \
  and\ \bibinfo {author} {\bibfnamefont {L.}~\bibnamefont {Juergen}},\ }\href
  {\doibase https://doi.org/10.1098/rsos.200780} {\bibfield  {journal}
  {\bibinfo  {journal} {R. Soc. open sci.}\ }\textbf {\bibinfo {volume} {7}},\
  \bibinfo {pages} {200780} (\bibinfo {year} {2020})}\BibitemShut {NoStop}%
\bibitem [{\citenamefont {Zhou}\ \emph {et~al.}(2020)\citenamefont {Zhou},
  \citenamefont {Liu}, \citenamefont {Yang}, \citenamefont {Liao},
  \citenamefont {Yang}, \citenamefont {Bai},\ and\ \citenamefont
  {Zhang}}]{Zhou2020}%
  \BibitemOpen
  \bibfield  {author} {\bibinfo {author} {\bibfnamefont {T.}~\bibnamefont
  {Zhou}}, \bibinfo {author} {\bibfnamefont {Q.}~\bibnamefont {Liu}}, \bibinfo
  {author} {\bibfnamefont {Z.}~\bibnamefont {Yang}}, \bibinfo {author}
  {\bibfnamefont {J.}~\bibnamefont {Liao}}, \bibinfo {author} {\bibfnamefont
  {K.}~\bibnamefont {Yang}}, \bibinfo {author} {\bibfnamefont {W.}~\bibnamefont
  {Bai}}, \ and\ \bibinfo {author} {\bibfnamefont {W.}~\bibnamefont {Zhang}},\
  }\href {\doibase https://doi.org/10.1098/rspa.1927.0118} {\bibfield
  {journal} {\bibinfo  {journal} {J Evid. Based Med.}\ }\textbf {\bibinfo
  {volume} {13}},\ \bibinfo {pages} {3} (\bibinfo {year} {2020})}\BibitemShut
  {NoStop}%
\bibitem [{\citenamefont {Prem}\ \emph {et~al.}(2020)\citenamefont {Prem},
  \citenamefont {Liu}, \citenamefont {Russel}, \citenamefont {Kucharski},
  \citenamefont {Eggo},\ and\ \citenamefont {Davies}}]{SEIR}%
  \BibitemOpen
  \bibfield  {author} {\bibinfo {author} {\bibfnamefont {K.}~\bibnamefont
  {Prem}}, \bibinfo {author} {\bibfnamefont {Y.}~\bibnamefont {Liu}}, \bibinfo
  {author} {\bibfnamefont {T.~W.}\ \bibnamefont {Russel}}, \bibinfo {author}
  {\bibfnamefont {A.~J.}\ \bibnamefont {Kucharski}}, \bibinfo {author}
  {\bibfnamefont {R.}~\bibnamefont {Eggo}}, \ and\ \bibinfo {author}
  {\bibfnamefont {N.}~\bibnamefont {Davies}},\ }\href {\doibase
  https://doi.org/10.1016/S2468-2667(20)30073-6} {\bibfield  {journal}
  {\bibinfo  {journal} {The Lancet Public Health}\ }\textbf {\bibinfo {volume}
  {5, issue 5}},\ \bibinfo {pages} {E261 } (\bibinfo {year}
  {2020})}\BibitemShut {NoStop}%
\bibitem [{\citenamefont {Scala}\ \emph {et~al.}(2020)\citenamefont {Scala},
  \citenamefont {Flori}, \citenamefont {Spelta}, \citenamefont {Brugnoli},
  \citenamefont {Cinelli}, \citenamefont {Quattrociocchi},\ and\ \citenamefont
  {Pammolli}}]{scala2020}%
  \BibitemOpen
  \bibfield  {author} {\bibinfo {author} {\bibfnamefont {A.}~\bibnamefont
  {Scala}}, \bibinfo {author} {\bibfnamefont {A.}~\bibnamefont {Flori}},
  \bibinfo {author} {\bibfnamefont {A.}~\bibnamefont {Spelta}}, \bibinfo
  {author} {\bibfnamefont {E.}~\bibnamefont {Brugnoli}}, \bibinfo {author}
  {\bibfnamefont {M.}~\bibnamefont {Cinelli}}, \bibinfo {author} {\bibfnamefont
  {W.}~\bibnamefont {Quattrociocchi}}, \ and\ \bibinfo {author} {\bibfnamefont
  {F.}~\bibnamefont {Pammolli}},\ }\href {\doibase
  https://doi.org/10.1038/s41598-020-70631-9} {\bibfield  {journal} {\bibinfo
  {journal} {Sci Rep}\ }\textbf {\bibinfo {volume} {10}},\ \bibinfo {pages}
  {13764} (\bibinfo {year} {2020})}\BibitemShut {NoStop}%
\bibitem [{\citenamefont {Friston}\ \emph {et~al.}(2020)\citenamefont
  {Friston}, \citenamefont {Parr}, \citenamefont {Zeidman}, \citenamefont
  {Razi}, \citenamefont {Flandin}, \citenamefont {Daunizeau}, \citenamefont
  {Hulme}, \citenamefont {Billig}, \citenamefont {Litvak}, \citenamefont
  {Price}, \citenamefont {Moran},\ and\ \citenamefont
  {Lambert}}]{friston2020second}%
  \BibitemOpen
  \bibfield  {author} {\bibinfo {author} {\bibfnamefont {K.~J.}\ \bibnamefont
  {Friston}}, \bibinfo {author} {\bibfnamefont {T.}~\bibnamefont {Parr}},
  \bibinfo {author} {\bibfnamefont {P.}~\bibnamefont {Zeidman}}, \bibinfo
  {author} {\bibfnamefont {A.}~\bibnamefont {Razi}}, \bibinfo {author}
  {\bibfnamefont {G.}~\bibnamefont {Flandin}}, \bibinfo {author} {\bibfnamefont
  {J.}~\bibnamefont {Daunizeau}}, \bibinfo {author} {\bibfnamefont {O.~J.}\
  \bibnamefont {Hulme}}, \bibinfo {author} {\bibfnamefont {A.~J.}\ \bibnamefont
  {Billig}}, \bibinfo {author} {\bibfnamefont {V.}~\bibnamefont {Litvak}},
  \bibinfo {author} {\bibfnamefont {C.~J.}\ \bibnamefont {Price}}, \bibinfo
  {author} {\bibfnamefont {R.~J.}\ \bibnamefont {Moran}}, \ and\ \bibinfo
  {author} {\bibfnamefont {C.}~\bibnamefont {Lambert}},\ }\href@noop {}
  {\enquote {\bibinfo {title} {Second waves, social distancing, and the spread
  of covid-19 across america},}\ } (\bibinfo {year} {2020}),\ \Eprint
  {http://arxiv.org/abs/2004.13017} {arXiv:2004.13017 [q-bio.PE]} \BibitemShut
  {NoStop}%
\bibitem [{\citenamefont {Sonnino}\ \emph {et~al.}(2020)\citenamefont
  {Sonnino}, \citenamefont {Mora},\ and\ \citenamefont
  {Nardone}}]{sonnino2020stochastic}%
  \BibitemOpen
  \bibfield  {author} {\bibinfo {author} {\bibfnamefont {G.}~\bibnamefont
  {Sonnino}}, \bibinfo {author} {\bibfnamefont {F.}~\bibnamefont {Mora}}, \
  and\ \bibinfo {author} {\bibfnamefont {P.}~\bibnamefont {Nardone}},\
  }\href@noop {} {\enquote {\bibinfo {title} {A stochastic compartmental model
  for covid-19},}\ } (\bibinfo {year} {2020}),\ \Eprint
  {http://arxiv.org/abs/2012.01869} {arXiv:2012.01869 [physics.med-ph]}
  \BibitemShut {NoStop}%
\bibitem [{\citenamefont {Abou-Ismail}(2020)}]{Anas2020}%
  \BibitemOpen
  \bibfield  {author} {\bibinfo {author} {\bibfnamefont {A.}~\bibnamefont
  {Abou-Ismail}},\ }\href {\doibase https://doi.org/
  10.1007/s42399-020-00330-z} {\bibfield  {journal} {\bibinfo  {journal} {SN
  Compr Clin Med.}\ ,\ \bibinfo {pages} {1}} (\bibinfo {year}
  {2020})}\BibitemShut {NoStop}%
\bibitem [{\citenamefont {Kermack}\ \emph {et~al.}(1927)\citenamefont
  {Kermack}, \citenamefont {McKendrick},\ and\ \citenamefont
  {Walker}}]{Kermack:1927}%
  \BibitemOpen
  \bibfield  {author} {\bibinfo {author} {\bibfnamefont {W.~O.}\ \bibnamefont
  {Kermack}}, \bibinfo {author} {\bibfnamefont {A.}~\bibnamefont {McKendrick}},
  \ and\ \bibinfo {author} {\bibfnamefont {G.~T.}\ \bibnamefont {Walker}},\
  }\href {\doibase https://doi.org/10.1098/rspa.1927.0118} {\bibfield
  {journal} {\bibinfo  {journal} {Proceedings of the Royal Society A}\ }\textbf
  {\bibinfo {volume} {115}},\ \bibinfo {pages} {700} (\bibinfo {year}
  {1927})}\BibitemShut {NoStop}%
\bibitem [{\citenamefont {Zhan}\ \emph {et~al.}(2018)\citenamefont {Zhan},
  \citenamefont {Liu}, \citenamefont {Zhou}, \citenamefont {Zhang},
  \citenamefont {Sun}, \citenamefont {Zhu},\ and\ \citenamefont
  {Jin}}]{ZHAN2018437}%
  \BibitemOpen
  \bibfield  {author} {\bibinfo {author} {\bibfnamefont {X.-X.}\ \bibnamefont
  {Zhan}}, \bibinfo {author} {\bibfnamefont {C.}~\bibnamefont {Liu}}, \bibinfo
  {author} {\bibfnamefont {G.}~\bibnamefont {Zhou}}, \bibinfo {author}
  {\bibfnamefont {Z.-K.}\ \bibnamefont {Zhang}}, \bibinfo {author}
  {\bibfnamefont {G.-Q.}\ \bibnamefont {Sun}}, \bibinfo {author} {\bibfnamefont
  {J.~J.}\ \bibnamefont {Zhu}}, \ and\ \bibinfo {author} {\bibfnamefont
  {Z.}~\bibnamefont {Jin}},\ }\href {\doibase
  https://doi.org/10.1016/j.amc.2018.03.050} {\bibfield  {journal} {\bibinfo
  {journal} {Applied Mathematics and Computation}\ }\textbf {\bibinfo {volume}
  {332}},\ \bibinfo {pages} {437 } (\bibinfo {year} {2018})}\BibitemShut
  {NoStop}%
\bibitem [{\citenamefont {Perc}\ \emph {et~al.}(2017)\citenamefont {Perc},
  \citenamefont {Jordan}, \citenamefont {Rand}, \citenamefont {Wang},
  \citenamefont {Boccaletti},\ and\ \citenamefont {Szolnoki}}]{PERC20171}%
  \BibitemOpen
  \bibfield  {author} {\bibinfo {author} {\bibfnamefont {M.}~\bibnamefont
  {Perc}}, \bibinfo {author} {\bibfnamefont {J.~J.}\ \bibnamefont {Jordan}},
  \bibinfo {author} {\bibfnamefont {D.~G.}\ \bibnamefont {Rand}}, \bibinfo
  {author} {\bibfnamefont {Z.}~\bibnamefont {Wang}}, \bibinfo {author}
  {\bibfnamefont {S.}~\bibnamefont {Boccaletti}}, \ and\ \bibinfo {author}
  {\bibfnamefont {A.}~\bibnamefont {Szolnoki}},\ }\href {\doibase
  https://doi.org/10.1016/j.physrep.2017.05.004} {\bibfield  {journal}
  {\bibinfo  {journal} {Physics Reports}\ }\textbf {\bibinfo {volume} {687}},\
  \bibinfo {pages} {1 } (\bibinfo {year} {2017})}\BibitemShut {NoStop}%
\bibitem [{\citenamefont {Wang}\ \emph {et~al.}(2015)\citenamefont {Wang},
  \citenamefont {Andrews}, \citenamefont {Wu}, \citenamefont {Wang},\ and\
  \citenamefont {Bauch}}]{WANG20151}%
  \BibitemOpen
  \bibfield  {author} {\bibinfo {author} {\bibfnamefont {Z.}~\bibnamefont
  {Wang}}, \bibinfo {author} {\bibfnamefont {M.~A.}\ \bibnamefont {Andrews}},
  \bibinfo {author} {\bibfnamefont {Z.-X.}\ \bibnamefont {Wu}}, \bibinfo
  {author} {\bibfnamefont {L.}~\bibnamefont {Wang}}, \ and\ \bibinfo {author}
  {\bibfnamefont {C.~T.}\ \bibnamefont {Bauch}},\ }\href {\doibase
  https://doi.org/10.1016/j.plrev.2015.07.006} {\bibfield  {journal} {\bibinfo
  {journal} {Physics of Life Reviews}\ }\textbf {\bibinfo {volume} {15}},\
  \bibinfo {pages} {1 } (\bibinfo {year} {2015})}\BibitemShut {NoStop}%
\bibitem [{\citenamefont {Scudellari}(2020)}]{Scudellari}%
  \BibitemOpen
  \bibfield  {author} {\bibinfo {author} {\bibfnamefont {M.}~\bibnamefont
  {Scudellari}},\ }\href {\doibase https://doi: 10.1038/d41586-020-02278-5}
  {\bibfield  {journal} {\bibinfo  {journal} {Nature}\ }\textbf {\bibinfo
  {volume} {584}},\ \bibinfo {pages} {22 } (\bibinfo {year}
  {2020})}\BibitemShut {NoStop}%
\bibitem [{\citenamefont {Della~Morte}\ \emph {et~al.}(2020)\citenamefont
  {Della~Morte}, \citenamefont {Orlando},\ and\ \citenamefont
  {Sannino}}]{DellaMorte:2020wlc}%
  \BibitemOpen
  \bibfield  {author} {\bibinfo {author} {\bibfnamefont {M.}~\bibnamefont
  {Della~Morte}}, \bibinfo {author} {\bibfnamefont {D.}~\bibnamefont
  {Orlando}}, \ and\ \bibinfo {author} {\bibfnamefont {F.}~\bibnamefont
  {Sannino}},\ }\href {\doibase https://doi.org/10.3389/fphy.2020.00144}
  {\bibfield  {journal} {\bibinfo  {journal} {Front. in Phys.}\ }\textbf
  {\bibinfo {volume} {8}},\ \bibinfo {pages} {144} (\bibinfo {year}
  {2020})}\BibitemShut {NoStop}%
\bibitem [{\citenamefont {Cacciapaglia}\ and\ \citenamefont
  {Sannino}(2020{\natexlab{a}})}]{Cacciapaglia:2020mjf}%
  \BibitemOpen
  \bibfield  {author} {\bibinfo {author} {\bibfnamefont {G.}~\bibnamefont
  {Cacciapaglia}}\ and\ \bibinfo {author} {\bibfnamefont {F.}~\bibnamefont
  {Sannino}},\ }\href {\doibase https://doi.org/10.1038/s41598-020-72175-4}
  {\bibfield  {journal} {\bibinfo  {journal} {Sci Rep}\ }\textbf {\bibinfo
  {volume} {10}},\ \bibinfo {pages} {15828} (\bibinfo {year}
  {2020}{\natexlab{a}})},\ \Eprint {http://arxiv.org/abs/2005.04956}
  {arXiv:2005.04956 [physics.soc-ph]} \BibitemShut {NoStop}%
\bibitem [{\citenamefont {Della~Morte}\ and\ \citenamefont
  {Sannino}(2020)}]{DellaMorte:2020qry}%
  \BibitemOpen
  \bibfield  {author} {\bibinfo {author} {\bibfnamefont {M.}~\bibnamefont
  {Della~Morte}}\ and\ \bibinfo {author} {\bibfnamefont {F.}~\bibnamefont
  {Sannino}},\ }\href@noop {} {\enquote {\bibinfo {title} {{Renormalisation
  Group approach to pandemics as a time-dependent SIR model}},}\ } (\bibinfo
  {year} {2020}),\ \Eprint {http://arxiv.org/abs/2007.11296} {arXiv:2007.11296
  [physics.soc-ph]} \BibitemShut {NoStop}%
\bibitem [{\citenamefont {Cacciapaglia}\ and\ \citenamefont
  {Sannino}(2020{\natexlab{b}})}]{cacciapaglia2020evidence}%
  \BibitemOpen
  \bibfield  {author} {\bibinfo {author} {\bibfnamefont {G.}~\bibnamefont
  {Cacciapaglia}}\ and\ \bibinfo {author} {\bibfnamefont {F.}~\bibnamefont
  {Sannino}},\ }\href {\doibase https://doi.org/10.21203/rs.3.rs-70238/v1}
  {\enquote {\bibinfo {title} {Evidence for complex fixed points in pandemic
  data},}\ } (\bibinfo {year} {2020}{\natexlab{b}}),\ \Eprint
  {http://arxiv.org/abs/2009.08861} {arXiv:2009.08861 [physics.soc-ph]}
  \BibitemShut {NoStop}%
\bibitem [{\citenamefont {Cacciapaglia}\ \emph
  {et~al.}(2020{\natexlab{a}})\citenamefont {Cacciapaglia}, \citenamefont
  {Cot},\ and\ \citenamefont {Sannino}}]{cacciapaglia2020multiwave}%
  \BibitemOpen
  \bibfield  {author} {\bibinfo {author} {\bibfnamefont {G.}~\bibnamefont
  {Cacciapaglia}}, \bibinfo {author} {\bibfnamefont {C.}~\bibnamefont {Cot}}, \
  and\ \bibinfo {author} {\bibfnamefont {F.}~\bibnamefont {Sannino}},\
  }\href@noop {} {\enquote {\bibinfo {title} {Multiwave pandemic dynamics
  explained: How to tame the next wave of infectious diseases},}\ } (\bibinfo
  {year} {2020}{\natexlab{a}}),\ \Eprint {http://arxiv.org/abs/2011.12846}
  {arXiv:2011.12846 [physics.soc-ph]} \BibitemShut {NoStop}%
\bibitem [{\citenamefont {Taubenberger}\ and\ \citenamefont
  {Morens}(2006)}]{1918influenza}%
  \BibitemOpen
  \bibfield  {author} {\bibinfo {author} {\bibfnamefont {J.~K.}\ \bibnamefont
  {Taubenberger}}\ and\ \bibinfo {author} {\bibfnamefont {D.~M.}\ \bibnamefont
  {Morens}},\ }\href@noop {} {\bibfield  {journal} {\bibinfo  {journal} {Rev
  Biomed}\ }\textbf {\bibinfo {volume} {17(1)}},\ \bibinfo {pages} {69}
  (\bibinfo {year} {2006})}\BibitemShut {NoStop}%
\bibitem [{\citenamefont {Wilson}(1971{\natexlab{a}})}]{Wilson:1971bg}%
  \BibitemOpen
  \bibfield  {author} {\bibinfo {author} {\bibfnamefont {K.~G.}\ \bibnamefont
  {Wilson}},\ }\href {\doibase https://doi.org/10.1103/PhysRevB.4.3174}
  {\bibfield  {journal} {\bibinfo  {journal} {Phys. Rev. B}\ }\textbf {\bibinfo
  {volume} {4}},\ \bibinfo {pages} {3174} (\bibinfo {year}
  {1971}{\natexlab{a}})}\BibitemShut {NoStop}%
\bibitem [{\citenamefont {Wilson}(1971{\natexlab{b}})}]{Wilson:1971dh}%
  \BibitemOpen
  \bibfield  {author} {\bibinfo {author} {\bibfnamefont {K.~G.}\ \bibnamefont
  {Wilson}},\ }\href {\doibase https://doi.org/10.1103/PhysRevB.4.3184}
  {\bibfield  {journal} {\bibinfo  {journal} {Phys. Rev. B}\ }\textbf {\bibinfo
  {volume} {4}},\ \bibinfo {pages} {3184} (\bibinfo {year}
  {1971}{\natexlab{b}})}\BibitemShut {NoStop}%
\bibitem [{\citenamefont {Li}\ \emph {et~al.}(2019)\citenamefont {Li},
  \citenamefont {Zhang}, \citenamefont {Liu}, \citenamefont {Zhang},
  \citenamefont {Wang},\ and\ \citenamefont {Wang}}]{LI2019566}%
  \BibitemOpen
  \bibfield  {author} {\bibinfo {author} {\bibfnamefont {L.}~\bibnamefont
  {Li}}, \bibinfo {author} {\bibfnamefont {J.}~\bibnamefont {Zhang}}, \bibinfo
  {author} {\bibfnamefont {C.}~\bibnamefont {Liu}}, \bibinfo {author}
  {\bibfnamefont {H.-T.}\ \bibnamefont {Zhang}}, \bibinfo {author}
  {\bibfnamefont {Y.}~\bibnamefont {Wang}}, \ and\ \bibinfo {author}
  {\bibfnamefont {Z.}~\bibnamefont {Wang}},\ }\href {\doibase
  https://doi.org/10.1016/j.amc.2018.11.042} {\bibfield  {journal} {\bibinfo
  {journal} {Applied Mathematics and Computation}\ }\textbf {\bibinfo {volume}
  {347}},\ \bibinfo {pages} {566 } (\bibinfo {year} {2019})}\BibitemShut
  {NoStop}%
\bibitem [{\citenamefont {Wang}\ \emph {et~al.}(2016)\citenamefont {Wang},
  \citenamefont {Bauch}, \citenamefont {Bhattacharyya}, \citenamefont
  {d'Onofrio}, \citenamefont {Manfredi}, \citenamefont {Perc}, \citenamefont
  {Perra}, \citenamefont {Salath\'{e}},\ and\ \citenamefont
  {Zhao}}]{WANG20161}%
  \BibitemOpen
  \bibfield  {author} {\bibinfo {author} {\bibfnamefont {Z.}~\bibnamefont
  {Wang}}, \bibinfo {author} {\bibfnamefont {C.~T.}\ \bibnamefont {Bauch}},
  \bibinfo {author} {\bibfnamefont {S.}~\bibnamefont {Bhattacharyya}}, \bibinfo
  {author} {\bibfnamefont {A.}~\bibnamefont {d'Onofrio}}, \bibinfo {author}
  {\bibfnamefont {P.}~\bibnamefont {Manfredi}}, \bibinfo {author}
  {\bibfnamefont {M.}~\bibnamefont {Perc}}, \bibinfo {author} {\bibfnamefont
  {N.}~\bibnamefont {Perra}}, \bibinfo {author} {\bibfnamefont
  {M.}~\bibnamefont {Salath\'{e}}}, \ and\ \bibinfo {author} {\bibfnamefont
  {D.}~\bibnamefont {Zhao}},\ }\href {\doibase
  https://doi.org/10.1016/j.physrep.2016.10.006} {\bibfield  {journal}
  {\bibinfo  {journal} {Physics Reports}\ }\textbf {\bibinfo {volume} {664}},\
  \bibinfo {pages} {1 } (\bibinfo {year} {2016})}\BibitemShut {NoStop}%
\bibitem [{\citenamefont {Danby}(1985)}]{Danby85}%
  \BibitemOpen
  \bibfield  {author} {\bibinfo {author} {\bibfnamefont {J.~M.~A.}\
  \bibnamefont {Danby}},\ }\href@noop {} {\emph {\bibinfo {title} {Computing
  applications to differential equations modelling in the physical and social
  sciences}}}\ (\bibinfo  {publisher} {Reston Publishing Company},\ \bibinfo
  {address} {Reston VA (USA)},\ \bibinfo {year} {1985})\BibitemShut {NoStop}%
\bibitem [{\citenamefont {Brauer}(2019)}]{Brauer2019}%
  \BibitemOpen
  \bibfield  {author} {\bibinfo {author} {\bibfnamefont {F.}~\bibnamefont
  {Brauer}},\ }\href {\doibase https://doi.org/10.1080/17513758.2018.1469792}
  {\bibfield  {journal} {\bibinfo  {journal} {Journal of Biological Dynamics}\
  }\textbf {\bibinfo {volume} {13}},\ \bibinfo {pages} {23} (\bibinfo {year}
  {2019})}\BibitemShut {NoStop}%
\bibitem [{\citenamefont {Miller}(2012)}]{Miller2012}%
  \BibitemOpen
  \bibfield  {author} {\bibinfo {author} {\bibfnamefont {J.~C.}\ \bibnamefont
  {Miller}},\ }\href {\doibase https://doi.org/10.1007/s11538-012-9749-6}
  {\bibfield  {journal} {\bibinfo  {journal} {Bulletin of mathematical
  biology}\ }\textbf {\bibinfo {volume} {74}},\ \bibinfo {pages} {2125}
  (\bibinfo {year} {2012})}\BibitemShut {NoStop}%
\bibitem [{\citenamefont {Murray}(2002)}]{Murray}%
  \BibitemOpen
  \bibfield  {author} {\bibinfo {author} {\bibfnamefont {J.~D.}\ \bibnamefont
  {Murray}},\ }\href@noop {} {\emph {\bibinfo {title} {Mathematical
  biology}}},\ \bibinfo {edition} {3rd}\ ed.,\ Interdisciplinary applied
  mathematics\ (\bibinfo  {publisher} {Springer},\ \bibinfo {address} {New York
  (USA)},\ \bibinfo {year} {2002})\BibitemShut {NoStop}%
\bibitem [{\citenamefont {Fishman}\ \emph {et~al.}(2014)\citenamefont
  {Fishman}, \citenamefont {Khoo},\ and\ \citenamefont {Tuite}}]{Fishman2014}%
  \BibitemOpen
  \bibfield  {author} {\bibinfo {author} {\bibfnamefont {D.}~\bibnamefont
  {Fishman}}, \bibinfo {author} {\bibfnamefont {E.}~\bibnamefont {Khoo}}, \
  and\ \bibinfo {author} {\bibfnamefont {A.}~\bibnamefont {Tuite}},\
  }\href@noop {} {\bibfield  {journal} {\bibinfo  {journal} {PLOS Currents
  Outbreaks}\ }\textbf {\bibinfo {volume} {6}} (\bibinfo {year}
  {2014})}\BibitemShut {NoStop}%
\bibitem [{\citenamefont {Pell}\ \emph {et~al.}(2018)\citenamefont {Pell},
  \citenamefont {Kuang}, \citenamefont {Viboud},\ and\ \citenamefont
  {Chowell}}]{Pell2018}%
  \BibitemOpen
  \bibfield  {author} {\bibinfo {author} {\bibfnamefont {B.}~\bibnamefont
  {Pell}}, \bibinfo {author} {\bibfnamefont {Y.}~\bibnamefont {Kuang}},
  \bibinfo {author} {\bibfnamefont {C.}~\bibnamefont {Viboud}}, \ and\ \bibinfo
  {author} {\bibfnamefont {G.}~\bibnamefont {Chowell}},\ }\href {\doibase
  https://doi.org/10.1016/j.epidem.2016.11.002} {\bibfield  {journal} {\bibinfo
   {journal} {Epidemics}\ }\textbf {\bibinfo {volume} {22}},\ \bibinfo {pages}
  {62 } (\bibinfo {year} {2018})},\ \bibinfo {note} {the RAPIDD Ebola
  Forecasting Challenge}\BibitemShut {NoStop}%
\bibitem [{\citenamefont {Paltiel}\ \emph {et~al.}(2020)\citenamefont
  {Paltiel}, \citenamefont {Schwartz}, \citenamefont {Zheng},\ and\
  \citenamefont {Walensky}}]{vaccine}%
  \BibitemOpen
  \bibfield  {author} {\bibinfo {author} {\bibfnamefont {A.~D.}\ \bibnamefont
  {Paltiel}}, \bibinfo {author} {\bibfnamefont {J.~L.}\ \bibnamefont
  {Schwartz}}, \bibinfo {author} {\bibfnamefont {A.}~\bibnamefont {Zheng}}, \
  and\ \bibinfo {author} {\bibfnamefont {R.~P.}\ \bibnamefont {Walensky}},\
  }\href@noop {} {\bibfield  {journal} {\bibinfo  {journal} {Health Affairs}\ }
  (\bibinfo {year} {2020})}\BibitemShut {NoStop}%
\bibitem [{\citenamefont {Sch{\"a}fer}\ \emph {et~al.}(2014)\citenamefont
  {Sch{\"a}fer}, \citenamefont {Strohmeier}, \citenamefont {Lenders},
  \citenamefont {Martinovic},\ and\ \citenamefont
  {Wilhelm}}]{schafer2014bringing}%
  \BibitemOpen
  \bibfield  {author} {\bibinfo {author} {\bibfnamefont {M.}~\bibnamefont
  {Sch{\"a}fer}}, \bibinfo {author} {\bibfnamefont {M.}~\bibnamefont
  {Strohmeier}}, \bibinfo {author} {\bibfnamefont {V.}~\bibnamefont {Lenders}},
  \bibinfo {author} {\bibfnamefont {I.}~\bibnamefont {Martinovic}}, \ and\
  \bibinfo {author} {\bibfnamefont {M.}~\bibnamefont {Wilhelm}},\ }in\
  \href@noop {} {\emph {\bibinfo {booktitle} {IPSN-14 Proceedings of the 13th
  International Symposium on Information Processing in Sensor Networks}}}\
  (\bibinfo {organization} {IEEE},\ \bibinfo {year} {2014})\ pp.\ \bibinfo
  {pages} {83--94}\BibitemShut {NoStop}%
\bibitem [{\citenamefont {Islind}\ \emph {et~al.}(2020)\citenamefont {Islind},
  \citenamefont {Óskarsdóttir},\ and\ \citenamefont
  {Steingrímsdóttir}}]{islind2020changes}%
  \BibitemOpen
  \bibfield  {author} {\bibinfo {author} {\bibfnamefont {A.~S.}\ \bibnamefont
  {Islind}}, \bibinfo {author} {\bibfnamefont {M.}~\bibnamefont
  {Óskarsdóttir}}, \ and\ \bibinfo {author} {\bibfnamefont {H.}~\bibnamefont
  {Steingrímsdóttir}},\ }\href@noop {} {\enquote {\bibinfo {title} {Changes
  in mobility patterns in europe during the covid-19 pandemic: Novel insights
  using open source data},}\ } (\bibinfo {year} {2020}),\ \Eprint
  {http://arxiv.org/abs/2008.10505} {arXiv:2008.10505 [cs.CY]} \BibitemShut
  {NoStop}%
\bibitem [{\citenamefont {{Bank of England}}(2020)}]{BankofEngland}%
  \BibitemOpen
  \bibfield  {author} {\bibinfo {author} {\bibnamefont {{Bank of England}}},\
  }\href@noop {} {\enquote {\bibinfo {title} {Monetary policy report and
  interim financial stability report - may 2020},}\ }\bibinfo {howpublished}
  {https://www.bankofengland.co.uk/report/2020/monetary-policy-report-financial-stability-report-may-2020}
  (\bibinfo {year} {2020})\BibitemShut {NoStop}%
\bibitem [{\citenamefont {Cacciapaglia}\ \emph
  {et~al.}(2020{\natexlab{b}})\citenamefont {Cacciapaglia}, \citenamefont
  {Cot},\ and\ \citenamefont {Sannino}}]{cacciapaglia2020second}%
  \BibitemOpen
  \bibfield  {author} {\bibinfo {author} {\bibfnamefont {G.}~\bibnamefont
  {Cacciapaglia}}, \bibinfo {author} {\bibfnamefont {C.}~\bibnamefont {Cot}}, \
  and\ \bibinfo {author} {\bibfnamefont {F.}~\bibnamefont {Sannino}},\ }\href
  {\doibase https://doi.org/10.1038/s41598-020-72611-5} {\bibfield  {journal}
  {\bibinfo  {journal} {Sci Rep}\ }\textbf {\bibinfo {volume} {10}},\ \bibinfo
  {pages} {15514} (\bibinfo {year} {2020}{\natexlab{b}})},\ \Eprint
  {http://arxiv.org/abs/2007.13100} {arXiv:2007.13100 [physics.soc-ph]}
  \BibitemShut {NoStop}%
\bibitem [{\citenamefont {Cardy}\ and\ \citenamefont
  {Grassberger}(1985)}]{Cardy_1985}%
  \BibitemOpen
  \bibfield  {author} {\bibinfo {author} {\bibfnamefont {J.~L.}\ \bibnamefont
  {Cardy}}\ and\ \bibinfo {author} {\bibfnamefont {P.}~\bibnamefont
  {Grassberger}},\ }\href {\doibase 10.1088/0305-4470/18/6/001} {\bibfield
  {journal} {\bibinfo  {journal} {Journal of Physics A: Mathematical and
  General}\ }\textbf {\bibinfo {volume} {18}},\ \bibinfo {pages} {L267}
  (\bibinfo {year} {1985})}\BibitemShut {NoStop}%
\bibitem [{\citenamefont {Yang}\ \emph {et~al.}(2020)\citenamefont {Yang},
  \citenamefont {Sha}, \citenamefont {Liu}, \citenamefont {Li}, \citenamefont
  {Lan}, \citenamefont {Guan}, \citenamefont {Hu}, \citenamefont {Li},
  \citenamefont {Zhang}, \citenamefont {Thompson},\ and\ \citenamefont
  {et~al.}}]{Yang_2020}%
  \BibitemOpen
  \bibfield  {author} {\bibinfo {author} {\bibfnamefont {C.}~\bibnamefont
  {Yang}}, \bibinfo {author} {\bibfnamefont {D.}~\bibnamefont {Sha}}, \bibinfo
  {author} {\bibfnamefont {Q.}~\bibnamefont {Liu}}, \bibinfo {author}
  {\bibfnamefont {Y.}~\bibnamefont {Li}}, \bibinfo {author} {\bibfnamefont
  {H.}~\bibnamefont {Lan}}, \bibinfo {author} {\bibfnamefont {W.~W.}\
  \bibnamefont {Guan}}, \bibinfo {author} {\bibfnamefont {T.}~\bibnamefont
  {Hu}}, \bibinfo {author} {\bibfnamefont {Z.}~\bibnamefont {Li}}, \bibinfo
  {author} {\bibfnamefont {Z.}~\bibnamefont {Zhang}}, \bibinfo {author}
  {\bibfnamefont {J.~H.}\ \bibnamefont {Thompson}}, \ and\ \bibinfo {author}
  {\bibnamefont {et~al.}},\ }\href {\doibase 10.1080/17538947.2020.1809723}
  {\bibfield  {journal} {\bibinfo  {journal} {International Journal of Digital
  Earth}\ }\textbf {\bibinfo {volume} {13}},\ \bibinfo {pages} {1186–1211}
  (\bibinfo {year} {2020})}\BibitemShut {NoStop}%
\bibitem [{\citenamefont {Priesemann}\ \emph {et~al.}(2020)\citenamefont
  {Priesemann}, \citenamefont {Brinkmann}, \citenamefont {Ciesek},
  \citenamefont {Cuschieri}, \citenamefont {Czypionka}, \citenamefont
  {Giordano}, \citenamefont {Gurdasani}, \citenamefont {Hanson}, \citenamefont
  {Hens}, \citenamefont {Iftekhar} \emph {et~al.}}]{priesemann2020calling}%
  \BibitemOpen
  \bibfield  {author} {\bibinfo {author} {\bibfnamefont {V.}~\bibnamefont
  {Priesemann}}, \bibinfo {author} {\bibfnamefont {M.~M.}\ \bibnamefont
  {Brinkmann}}, \bibinfo {author} {\bibfnamefont {S.}~\bibnamefont {Ciesek}},
  \bibinfo {author} {\bibfnamefont {S.}~\bibnamefont {Cuschieri}}, \bibinfo
  {author} {\bibfnamefont {T.}~\bibnamefont {Czypionka}}, \bibinfo {author}
  {\bibfnamefont {G.}~\bibnamefont {Giordano}}, \bibinfo {author}
  {\bibfnamefont {D.}~\bibnamefont {Gurdasani}}, \bibinfo {author}
  {\bibfnamefont {C.}~\bibnamefont {Hanson}}, \bibinfo {author} {\bibfnamefont
  {N.}~\bibnamefont {Hens}}, \bibinfo {author} {\bibfnamefont {E.}~\bibnamefont
  {Iftekhar}},  \emph {et~al.},\ }\href@noop {} {\bibfield  {journal} {\bibinfo
   {journal} {The Lancet}\ } (\bibinfo {year} {2020})}\BibitemShut {NoStop}%
\end{thebibliography}%

\section*{Author contribution}

This work has been designed and performed conjointly and equally by the authors, who have equally contributed to the writing of the article. ASI and MO have extracted and processed the data from flights; CC has worked on the numerical results from the eRG equations and analysed the epidemiological data.

\section*{Competing interests}

The authors declare no competing interests.

\end{document}